\documentclass[sigconf]{acmart}
\usepackage{algorithmic}
\usepackage{graphicx}
\usepackage{textcomp}
\usepackage{graphicx}
\usepackage{caption}
\usepackage{subcaption}
\usepackage{bbm}
\usepackage{color}
\usepackage{hyperref}
\usepackage{multirow}
\usepackage{enumitem}
\usepackage{algorithm} 
\usepackage{algorithmic}
\usepackage{enumerate}
\usepackage{amsmath}
\usepackage{booktabs}
\usepackage{multirow}
\usepackage{stmaryrd}
\usepackage{mathrsfs} 
\usepackage{color}
\usepackage{balance}
\usepackage{adjustbox}
\usepackage{tcolorbox}

\usepackage[para,online,flushleft]{threeparttable}

\newcommand{\eg}{\textit{e.g.}}

\newcommand{\ie}{\textit{i.e.}}

\AtBeginDocument{%
  }

\copyrightyear{2025}
\acmYear{2025}
\setcopyright{acmlicensed}\acmConference[SIGIR '25]{Proceedings of the 48th International ACM SIGIR Conference on Research and Development in Information Retrieval}{July 13--18, 2025}{Padua, Italy}
\acmBooktitle{Proceedings of the 48th International ACM SIGIR Conference on Research and Development in Information Retrieval (SIGIR '25), July 13--18, 2025, Padua, Italy}
\acmDOI{10.1145/3726302.3729961}
\acmISBN{979-8-4007-1592-1/2025/07}

\begin{document}

\title{Efficiency Unleashed: Inference Acceleration for LLM-based Recommender Systems with Speculative Decoding}

\author{Yunjia Xi}
\authornote{Both authors contributed equally to this research.}
\email{xiyunjia@sjtu.edu.cn}
\affiliation{%
  \institution{Shanghai Jiao Tong University}
  \city{Shanghai}
  \country{China}
}

\author{Hangyu Wang}
\authornotemark[1]
\email{hangyuwang@sjtu.edu.cn}
\affiliation{%
  \institution{Shanghai Jiao Tong University}
  \city{Shanghai}
  \country{China}
}

\author{Bo Chen}
\email{chenbo116@huawei.com}
\affiliation{%
  \institution{Huawei Noah's Ark Lab}
  \city{Shanghai}
  \country{China}
}

\author{Jianghao Lin}
\authornote{Corresponding authors.}
\email{chiangel@sjtu.edu.cn}
\affiliation{%
  \institution{Shanghai Jiao Tong University}
  \city{Shanghai}
  \country{China}
}

\author{Menghui Zhu}
\email{zhumenghui1@huawei.com}
\affiliation{%
  \institution{Huawei Noah's Ark Lab}
  \city{Shanghai}
  \country{China}
}

\author{Weiwen Liu}
\email{liuweiwen8@huawei.com}
\affiliation{%
  \institution{Huawei Noah's Ark Lab}
  \city{Shenzhen}
  \country{China}
}

\author{Ruiming Tang}
\email{tangruiming@huawei.com}
\affiliation{%
  \institution{Huawei Noah's Ark Lab}
  \city{Shenzhen}
  \country{China}
}

\author{Zhewei Wei}
\email{zhewei@ruc.edu.cn}
\affiliation{%
  \institution{Renmin University of China}
  \city{Beijing}
  \country{China}
}

\author{Weinan Zhang}
\email{wnzhang@sjtu.edu.cn}
\affiliation{%
  \institution{Shanghai Jiao Tong University}
  \city{Shanghai}
  \country{China}
}

\author{Yong Yu}
\email{yyu@sjtu.edu.cn}
\affiliation{%
  \institution{Shanghai Jiao Tong University}
  \city{Shanghai}
  \country{China}
}

\renewcommand{\shortauthors}{Yunjia Xi et al.}

\begin{abstract}
  The past few years have witnessed a growing interest in LLM-based recommender systems (RSs), although their industrial deployment remains in a preliminary stage. Most existing deployments leverage LLMs offline as feature enhancers, generating augmented knowledge for downstream tasks. However, in recommendation scenarios with numerous users and items, even offline knowledge generation with LLMs demands significant time and computational resources. This inefficiency arises from the autoregressive nature of LLMs. A promising solution is speculative decoding, a Draft-Then-Verify approach that increases the number of tokens generated per decoding step.
In this work, we first identify recommendation knowledge generation as a highly fitting use case for retrieval-based speculative decoding. Then, we discern its two characteristics: (1) the vast number of items and users in RSs leads to \textbf{retrieval inefficiency}, and (2) RSs exhibit high \textbf{diversity tolerance} for LLM-generated text. Building on these insights, we introduce \underline{L}ossless \underline{A}cceleration via \underline{S}peculative D\underline{e}coding for LLM-based \underline{R}ecommender Systems (\textit{LASER}), which features a \textit{Customized Retrieval Pool} to enhance retrieval efficiency and \textit{Relaxed Verification} to improve the acceptance rate of draft tokens. LASER achieves a \textbf{3-5x} speedup on public datasets and saves about \textbf{67\%} of computational resources during the online A/B test on a large-scale advertising scenario with lossless downstream recommendation performance. Our code is available at \url{https://github.com/YunjiaXi/LASER}
\end{abstract}




\maketitle

\vspace{-5pt}
\section{Introduction}
Large language models (LLMs) are revolutionizing numerous domains through their extensive  capabilities~\cite{bubeck2023sparks,zhao2023survey,gpt4,zhu2023large}. In recommender systems (RSs), integrating LLMs has emerged as a prominent research focus~\cite{yu2023self,lin2023can,li2023large,fan2023recommender,wu2023survey}. Commercial RSs typically need to process data pertaining to billions of users and items, necessitating low response latency, often within 100 milliseconds~\cite{xi2023device}.  However, LLMs' enormous parameters and considerable inference latency hinder their deployment into commercial RSs that demand rapid response. To handle this challenge, industrial solutions commonly involve deploying LLMs offline as feature enhancers~\cite{xi2023towards,liu2024modeling,liu2024once,ren2024representation}. First, LLMs leverage their reasoning capabilities and extensive knowledge to generate augmented knowledge for RSs -- such as user profiles or tags~\cite{li2023taggpt} and supplementary knowledge or summaries for items~\cite{lyu2023llm}. This newly generated knowledge is subsequently incorporated as additional features into traditional recommendation models via text encoder~\cite{xi2023towards,luo2024kellmrec} or converting to categorical features~\cite{brinkmann2023product,fang2024llm}. This strategy capitalizes on the extensive knowledge and sophisticated reasoning capabilities of LLMs while satisfying the response latency demands of commercial RSs.

Even when leveraging LLMs offline for knowledge generation, the recommendation scenarios, characterized by a vast number of users and items, still face significant time and resource constraints. LLMs inherently have low inference efficiency coupled with substantial resource demands. The numerous items and users in RSs require frequent invocations of LLMs, leading to considerable resource and time consumption.
Taking Qwen-7B-Chat~\cite{bai2023qwen} as an example, it requires 4.88s to generate a piece of user preference knowledge of 250 tokens on an NVIDIA V100, and generating knowledge for an industrial-scale quantity of users, say 10 million, would take roughly 565 GPU days. Furthermore, this knowledge generation is a continual process since user preferences may vary with their behaviors, necessitating knowledge updates. 
Moreover, prolonged overall generation times can lead to delays in generating knowledge for new items and user behaviors, thereby impairing recommendation effectiveness. High resource consumption and low inference efficiency have emerged as significant obstacles to deploying LLMs in RSs. Thus, improving inference efficiency has become critical for the effective deployment of LLMs in RSs.

One of the bottlenecks in LLM inference stems from autoregressive decoding, which demands forwarding through a billion-parameter LLM to produce just a single token at each decoding step, and these steps cannot be parallelized. Recently, a promising direction for accelerating LLMs is \textbf{speculative decoding}, a \textit{Draft-then-Verify} paradigm that increases the number of generated tokens per decoding step~\cite{xia2024unlocking,he2023rest}. At each decoding step, it first efficiently drafts multiple future tokens via auxiliary models or database retrieval and then verifies all these draft tokens in parallel with target LLMs to speed up inference~\cite{xia2024unlocking}. By allowing multiple tokens to be generated in a single decoding step, speculative decoding diminishes the total number of decoding steps, thus improving inference efficiency with lossless generation accuracy.

\begin{figure}[]
    \centering
    \vspace{-8pt}
    \includegraphics[clip,width=\columnwidth]{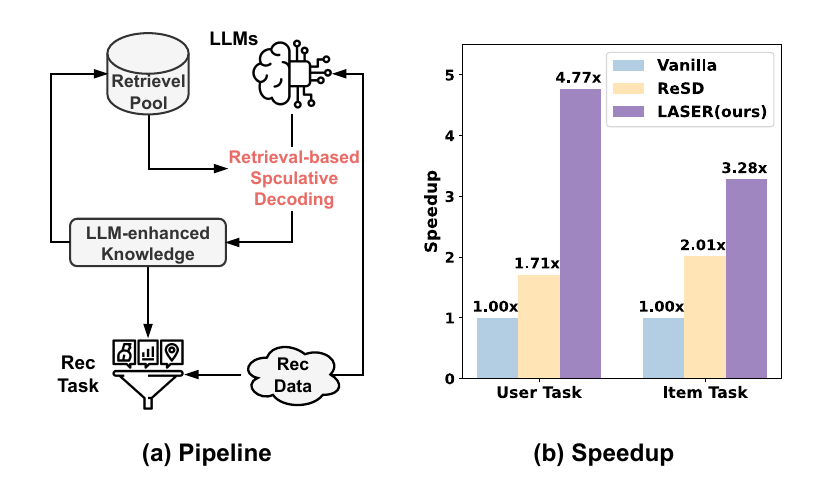}
    \vspace{-22pt}
    \caption{Pipeline of retrieval-based speculative decoding for RSs and speedup of autoregressive decoding (Vanilla), naive retrieval-based speculative decoding (ReSD), and LASER. 
    }
    \vspace{-15pt}
    \label{fig:intro}
\end{figure}

The knowledge generation based on LLMs in RSs exhibits specific properties that make it suitable for retrieval-based speculative decoding. \textit{Firstly,  recommendation knowledge generation is a continual process.} As user behaviors evolve and new items are introduced, we need to continuously generate new knowledge for new items and users' new behaviors, while we also possess much old knowledge about users' past behaviors and existing items. \textit{Secondly, there is often reusable content between new and old knowledge.} For instance, old and new user profiles may overlap due to user preference continuity. Therefore, we can utilize old knowledge as a retrieval pool to extract draft texts and then use LLMs to verify, thereby accelerating the generation of new knowledge, as shown in Figure~\ref{fig:intro}(a).

However, in practice, we find that this straightforward application overlooks some traits of RSs, leading to sub-optimal acceleration performance. 
Firstly, extensive items and users in RSs result in \textbf{retrieval inefficiency}, which impairs acceleration. A retrieval pool constructed with existing knowledge from all the users and items would be exceedingly large, which would significantly extend retrieval times. Therefore, it is essential to maintain smaller retrieval pools with similar knowledge, ensuring both low retrieval time and a high acceptance rate of draft tokens. Secondly, RSs exhibit high \textbf{diversity tolerance} for text generated by LLMs. Downstream recommendation tasks can achieve similar outcomes with texts that are not identical but have semantic proximity. In other words, RSs do not require perfectly consistent texts, which provides speculative decoding with further room for acceleration.

Based on the above insights, we propose a \underline{L}ossless \underline{A}cceleration via \underline{S}peculative D\underline{e}coding for LLM-based \underline{R}ecommender Systems (dubbed \textit{LASER}). We introduce two key enhancements to the retrieval-based speculative decoding for recommendation. 
\textit{Firstly}, \textbf{Customized Retrieval Pool} is designed to enhance retrieval efficiency. We introduce collaborative-based and attribute-based retrieval pool construction schemes, with a binary router to assign the appropriate retrieval pool to users and items. These personalized, compact retrieval pools maintain knowledge similarity, thereby guaranteeing low retrieval time and high acceptance rates of draft tokens. \textit{Next}, \textbf{Relaxed Verification} is devised to further enhance the acceptance rate of draft tokens. Traditional speculative decoding only accepts the token with the highest probability. We relax this restriction to top-$k$ probable tokens, increasing the number of accepted tokens while maintaining semantic proximity. Additionally, a probability threshold is imposed to prevent divergence during generation. The contributions of this work can be summarized as follows:
\begin{itemize}
    \item We identify the inefficiency of knowledge generation during deploying LLM-based recommendations and propose LASER. To the best of our knowledge, this is the first work to introduce \textbf{speculative decoding} into \textbf{LLM-based recommendations}, promoting the deployment of LLMs in RSs.
    \item We first discover two key traits of speculative decoding in RSs and implement two enhancements: \textbf{Customized Retrieval Pool} to improve retrieval efficiency and \textbf{Relaxed Verification} to increase accepted draft tokens.
    \item LASER achieves \textbf{3-5x speedup} and during online A/B test on a large-scale advertising scenario, it saves about \textbf{67\% of computational resources} with lossless recommendation performance.
\end{itemize}

\section{Preliminary Findings}\label{sect:preli}
\subsection{Speculative Decoding for Recommendation}
The mainstream Transformer-based LLMs typically adopt \textit{autoregressive decoding}. With the input token sequence $\{x_1,\ldots x_t\}$, the language model $\mathcal{M}$ generate next token following:
\begin{equation}
    x_{t+1}\sim q_{t+1} = \mathcal{M}(x|x_{\le t}),
\end{equation}
where $q_{t+1}$ denotes the conditional probability distribution from $\mathcal{M}$ and $x_{t+1}$ is the next token sampled from $q_{t+1}$. After this, $\mathcal{M}$ follows the same process to generate the next token. Despite desirable generation quality, autoregressive decoding only produces a single token per decoding step, making it inefficient and time-consuming.

To this end, \textit{speculative decoding}~\cite{cai2024medusa,he2023rest,zhao2023lookahead} have been proposed to generate a sequence of tokens at each decoding step. It is a Draft-then-Verify decoding paradigm in which, at each decoding step, it first efficiently drafts multiple future tokens and then verifies all these tokens in parallel with the target LLM~\cite{xia2024unlocking}. There are many strategy for draft generation, \eg, employing a small LM~\cite{leviathan2023fast}, retrieving from database~\cite{he2023rest,zhao2023lookahead}. Our work mainly focuses on retrieval-based draft models, which retrieve drafts from a given retrieval pool, since small LMs might lack recommendation knowledge and recommendation knowledge generation can provide appropriate retrieval pools naturally. Here, we take it as an example and delve into its two substeps -- drafting and verification.

\textbf{Drafting} phase is responsible for efficiently drafting multiple future tokens. Formally, given an input sequence $\{x_1,\ldots,x_t\}$, a draft model $\widetilde{\mathcal{M}}$ is employed, (\eg, a retriever that retrieves relevant text from the database) to generate the next $K$ draft tokens:
\begin{equation}
    \tilde{x}_1, \ldots, \tilde{x}_K = \text{Draft}(x_{\le t}, \widetilde{\mathcal{M}}),
\end{equation} 
where $\tilde{x}_i,i=1,\ldots,K$ denotes the drafted token generated by $\widetilde{\mathcal{M}}$ and $\text{Draft}(\cdot)$ represents draft generation strategies. 

\textbf{Verification} phase utilizes the target LLM to verify all these draft tokens in parallel. With the input sequence $\{x_1,\ldots,x_t\}$ and the draft sequence $\{\tilde{x_1}, \ldots, \tilde{x_K}\}$, the target LLM $\mathcal{M}$ calculates $K+1$ probability distributions simultaneously,
\begin{equation}
    q_i = \mathcal{M}(x_{\le t}, \tilde{x}_{<i}), i=1,\ldots,K+1.
\end{equation}
Then, each draft token $\tilde{x}_i$ is sequentially verified by a specific criterion $\text{Verify}(\tilde{x}_i, q_i)$. Typically, \textit{greedy verification} is adopted for retrieval-based speculative decoding via
\begin{equation}
    \tilde{x}_i = \arg \max q_i.
    \label{eq:GR}
\end{equation}
Only $\tilde{x}_i$ that meets the criterion in Eq~\eqref{eq:GR} is selected as final output, \ie, $x_{t+i}=\tilde{x}_i$. If a drafted token $\tilde{x}_c$ at position $c$ fails the verification, it will be corrected by distribution $q_c$ from target LLM, \ie, $x_{t+c}\leftarrow \arg\max q_c$. All drafted tokens after position $c$ will be discarded, ensuring quality consistent with the target LLM's standards.

The characteristics of recommendation knowledge generation make it highly suitable for applying retrieval-based speculative decoding (REST). REST requires a retrieval pool that overlaps with the currently generated text. As user behaviors evolve and new items are introduced, we need to continuously generate new knowledge for new items and users' new behaviors, resulting in a constant stream of old knowledge about users' past behaviors and existing items. Furthermore, there are notable similarities between this old knowledge and new knowledge. For instance, parts of old and new user profiles may overlap. Consequently, we can leverage old knowledge to construct retrieval pools and utilize REST to accelerate the generation of new knowledge.

\subsection{Finding 1: Retrieval Inefficiency}\label{sec:finding1}

According to the above approach, we conduct preliminary experiments on the MovieLens-10M dataset, following the setting of KAR~\cite{xi2023towards}, which first employs LLMs to generate recommendation knowledge and then adapts the knowledge to the downstream tasks. Here, Vicuna-7b-v1.3\footnote{\url{https://huggingface.co/lmsys/vicuna-7b-v1.3}} is leveraged to generate fine-grained user preferences based on user behaviors. To implement retrieval-based speculative decoding, we simulate users' streaming behaviors and divide the behaviors into multiple segments. For simplicity, the user's historical behavior $\{x_1,\ldots,x_n\}$ is divided into two segments: old history $x_1,\ldots,x_m$ and new history $x_{n-m},\ldots,x_n$, where $(\frac{n}{2}<m<n)$. Then, Vicuna-7b-v1.3 generates knowledge for all the old history with autoregressive decoding, based on which a retrieval pool is constructed. During knowledge generation for new history, we adopt speculative decoding, which retrieves drafts from the retrieval pool and uses Vicuna-7b-v1.3 to validate.

\begin{figure}[h]
    \centering
    \vspace{-10pt}
    \includegraphics[clip,width=\columnwidth]{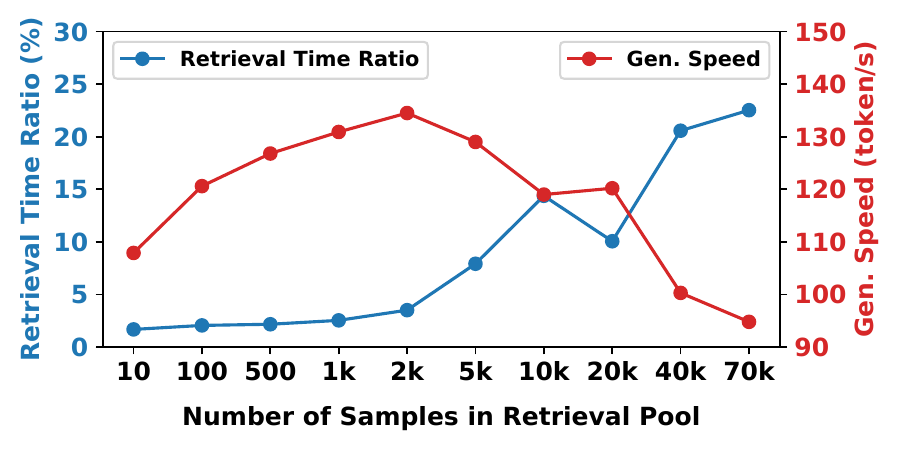}
    \vspace{-20pt}
    \caption{The impact of retrieval pool size.}
    \vspace{-8pt}
    \label{fig:size}
\end{figure}

Under the above conditions, we explore how the token generation speed (Gen. Speed) and the proportion of time spent retrieving relevant text to the total time (Retrieval Time Ratio) change when constructing the retrieval pool with different numbers of old knowledge samples (ranging from 10 to the maximum number of users) in Figure~\ref{fig:size}. In this figure, generation speed initially rises and then falls as the number of knowledge entries in the retrieval pool increases. When the retrieval pool is constructed with all users' old knowledge, the retrieval time ratio exceeds 20\%, causing generation speed to drop from its peak of 134.5 token/s to 94.8 token/s, significantly affecting the acceleration. When faced with an industrial-scale number of users, such as 10 million, the retrieval pool becomes larger, which will exacerbate the retrieval inefficiency.


Therefore, a large retrieval pool is not always advantageous. Although a larger retrieval pool can provide a greater volume of pertinent content, it also brings retrieval inefficiency, thereby impairing acceleration. It is imperative to construct an optimal retrieval pool that maintains low retrieval time while encompassing content similar to the text being generated.

\subsection{Finding 2: Diversity Tolerance}\label{sec:finding2}
Furthermore, we also investigate the impact of the diversity of LLM-generated texts on downstream tasks. Similar to the previous experiment, we leverage Vicuna-7b-v1.3 to generate user preference knowledge on MovieLens-10M. However, during generation, we sample from the top-$k$ most likely tokens to create approximate but diverse texts. We then adapt the encoding of knowledge from BERT to the CTR prediction task in RSs following~\cite{xi2023towards}. Specifically, we generate four different sets of user preference knowledge for all the users in the dataset. The knowledge is then applied to two well-known CTR models, DIN~\cite{DIN} and DCNv2~\cite{DCNv2}, with their performance in terms of AUC and Logloss presented in Table~\ref{tab:diversity}. In the table, \textbf{"w/o augment"} refers to results without knowledge augmentation, while \textbf{"knowledge 1"} to \textbf{"knowledge 4"} denotes results augmented with knowledge generated under different samplings.

\begin{table}[h]

    \vspace{-5pt}
    \caption{Performance comparison between CTR models augmented by different knowledge. }
    \vspace{-5pt}
    \centering
    \scalebox{1}{
    \setlength{\tabcolsep}{2.0mm}
    {\begin{tabular}{c|cc|cc}
\toprule
\multirow{2}{*}{Method} & \multicolumn{2}{c|}{DIN} & \multicolumn{2}{c}{DCNv2} \\
\cmidrule{2-5}
 & AUC & LL & AUC & LL \\
 \midrule
w/o augment & 0.8163 & 0.3619 & 0.8115 & 0.3663 \\
\midrule
knowledge 1 & 0.8351 & 0.3469 & 0.8319 & 0.3500 \\
knowledge 2 & 0.8353 & 0.3465 & 0.8314 & 0.3503 \\
knowledge 3 & 0.8347 & 0.3466 & 0.8319 & 0.3499 \\
knowledge 4 & 0.8349 & 0.3470 & 0.8323 & 0.3501
\\ 
\bottomrule
\end{tabular}}
}
\label{tab:diversity}
\vspace{-5pt}
\end{table}

The results in the table indicate that recommendation tasks exhibit a high diversity tolerance for LLM-generated knowledge texts. Compared to models without augmentation,  knowledge augmentation can result in a significant improvement, ranging from 1.5\% to 2\%. However, the performance difference between the diverse knowledge texts (knowledge 1-4) applied to downstream tasks is less than 0.1\%, showing that recommendation tasks are not sensitive to the diversity of LLM-generated texts.

Previously, retrieval-based speculative decoding typically adopts greedy verification, which only accepts the token with the highest probability to ensure text consistency with autoregressive decoding. However, this strict verification limits the acceptance rate of draft tokens. Given that downstream tasks in RSs can tolerate diverse LLM texts, we can consider relaxing the verification, allowing speculative decoding to accept more draft tokens and generate more diverse texts, thereby further enhancing the acceleration.

\section{Methodology}
\begin{figure*}
    \centering
    \includegraphics[clip,width=\textwidth]{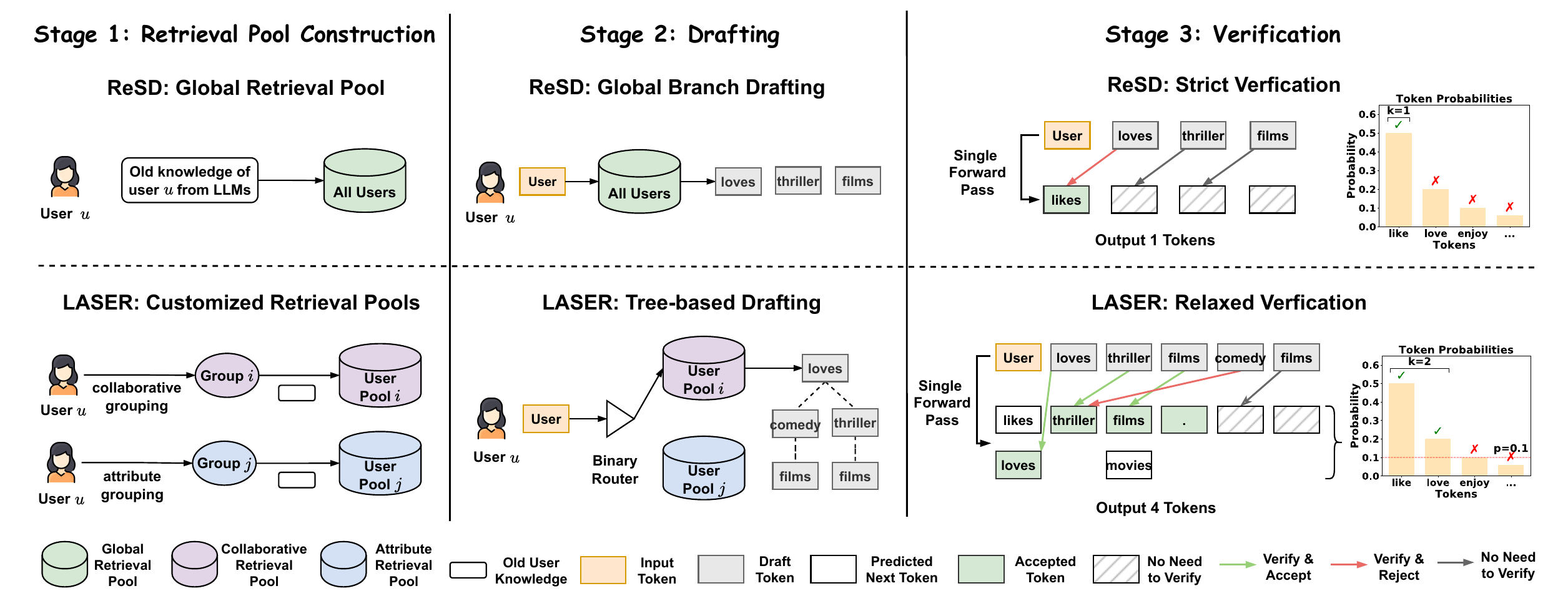}
    \vspace{-10pt}
    \caption{Comparison between naive retrieval-based speculative decoding ReSD (above), and our LASER (below). Here, we take users as examples, and the process is applicable to items. Note that the retrieved tree-structured draft is converted into a pseudo-sequence for parallel validation, which will be detailed in Section~\ref{sec:tree-based-draft}.}
    \label{fig:framework}
\end{figure*}
\subsection{Overview}
Based on the findings above, we have devised two enhancements for retrieval-based speculative decoding in recommendation knowledge generation. The \textbf{Customized Retrieval Pool} involves creating smaller retrieval pools tailored for similar items or users, thereby achieving low retrieval time. \textbf{Relaxed Verification} loosens the condition of greedy verification, which only accepts the highest probability token, to include the top-$k$ most likely tokens, thereby increasing the acceptance rate of draft tokens.

The workflow of our proposed LASER, illustrated in Figure~\ref{fig:framework}, encompasses three stages: customized retrieval pool construction, tree-based drafting, and relaxed verification. Before text generation, \textbf{customized retrieval pool construction} stage uses previously generated recommendation knowledge to build personalized retrieval pools in the form of trie tree~\cite{de1959file}. We first divide the users and items into different groups and then construct a retrieval pool for each group. The subsequent stage, \textbf{tree-based drafting}, retrieves relevant content from the designated retrieval pool when generating new knowledge for a specific user or item. This process yields a pseudo-sequence from a trie subtree that encapsulates multiple potential successor texts with an associated attention mask, subsequently validated in parallel by the target LLM. In the \textbf{relaxed verification} stage, we accept tokens from the top-$k$ highest-probability tokens that exceed a certain probability threshold $p$. This allows more draft tokens to be accepted and prevents divergence during generation, further improving the generation speed.
\vspace{-5pt}
\subsection{Customized Retrieval Pool Construction}
As mentioned in Section~\ref{sec:finding2}, a customized retrieval pool requires moderate capacity and internal knowledge similarity. This necessitates the partition of users and items, with distinct retrieval pools assigned to different groups. To maintain knowledge similarity, we can incorporate collaborative signals to group similar users and items. A common method involves clustering items or users based on their embeddings derived from recommendation models trained on user-item interactions. However, newly introduced users and items may have limited or no interaction records, making it challenging to obtain reliable embeddings. Considering that users or items with similar attributes may be more alike, their attributes can serve as a basis for constructing similar groups. Therefore, we design two retrieval pool construction schemes: one based on collaborative signals and the other on attributes. A binary router is devised to choose a retrieval pool for each user and item.

\textbf{Collaborative-based retrieval pool} groups items or users by clustering their embeddings containing collaborative signals. Initially, a recommendation model is trained on user-item interactions (\eg, LightGCN~\cite{he2020lightgcn}) and subsequently provides related embeddings, such as those of IDs and attributes. Given that RSs continuously train models for recommendations, we can re-utilize these embeddings. Then clustering algorithms, such as K-means~\cite{krishna1999genetic}, are applied to these embeddings to obtain distinct user or item groups. Users or items within the same cluster exhibit similarities, thereby ensuring that the knowledge generated by LLMs is more homogeneous. This, in turn, increases the probability of retrieving relevant texts.

\textbf{Attribute-based retrieval pool} partitions items or users by similar attributes, when well-trained embeddings are lacking. Items or users with similar attributes are more likely to exhibit higher similarity, resulting in more consistent knowledge generation by LLMs. Thus, items or users with analogous attributes, such as category, can be placed in the same group. If the sizes of groups formed based on general attributes like category exceed a certain threshold, we further subdivide them with additional attributes, \eg, subcategory, which is selected by manually crafted rules or decision trees.

Subsequently, for each item or user in a group, if there is previously generated knowledge from LLMs, this knowledge will be used to construct a retrieval pool for this group in the form of a trie tree~\cite{de1959file}. Trie tree is a data structure widely used for efficient retrieval and storage, as it efficiently handles prefix matching with each node as individual characters or words. Each group maintains its own trie tree, where each node represents a token and a path from the root node to a leaf node constitutes a branch~\cite{zhao2023lookahead}. These branches built with previously generated knowledge mentioned above are all \textbf{permanent branches} that would not be eliminated. During the generation of new knowledge, the current prompt and newly generated text are also relevant to subsequent generations. Therefore, we dynamically add the prompt and new content to the trie tree as \textbf{temporary branch} for the generation of each knowledge. As these additions may not necessarily enhance the acceleration of other knowledge generation, the branch will be eliminated once the generation is completed.

After constructing the collaborative-based and attribute-based retrieval pools, a \textbf{binary router} is designed to select the appropriate retrieval pool for users and items needing knowledge generation. It is highly flexible and can support various selection schemes. Based on our motivation, the default scheme is to select the collaborative-based retrieval pool for items and users with extensive interaction histories, while new users and items with few or no interactions are assigned to the attribute-based retrieval pool.

\subsection{Tree-based Drafting}\label{sec:tree-based-draft}
Before generating new knowledge for a user or item, we identify the corresponding retrieval pool $D$ based on their binary router, IDs, and attributes. During the knowledge generation, we first retrieve the relevant sub-tree from $D$ based on the current input text, which represents multiple potential successor sequences. Next, the pseudo-sequence, attention mask matrix, and position IDs for this sub-tree are generated to facilitate parallel validation by the target LLM with tree attention. Specifically, assuming the current input sequence is $\{x_1,\ldots,x_t\}$, the last $n$ tokens of the input sequence are adopted as a prefix to extract a sub-tree $T_t$ from $D$ as follows
\begin{equation}
    T_t = \text{Retrieve}(D, \{x_{t-n},\ldots, x_t\}, K) 
\end{equation}
where $\text{Retrieve}(\cdot)$ denotes retrieving a sub-tree from the trie tree with a prefix, and $K$ is the maximum length of draft tokens. The sub-tree $T_i$ is also a prefix tree, with each branch representing a potential successor draft sequence. Short prefixes yield a lot of content but may not be highly relevant, while long prefixes ensure high relevance but might fail to retrieve any content. Therefore, we will dynamically adjust $n$ during the retrieval process following~\cite{zhao2023lookahead}. Initially, a relatively large $n$, \ie, a long prefix, is used to guarantee relevance. If the number of retrieved tokens is significantly fewer than the maximum length $K$, we decrease $n$ to retry the retrieval process further until obtaining a substantial number of tokens. Conversely, if the number of retrieved tokens exceeds $K$, the tokens with the highest frequency are selected as draft tokens.

\begin{figure}[h]
    \centering
    \includegraphics[clip,width=1.06\columnwidth]{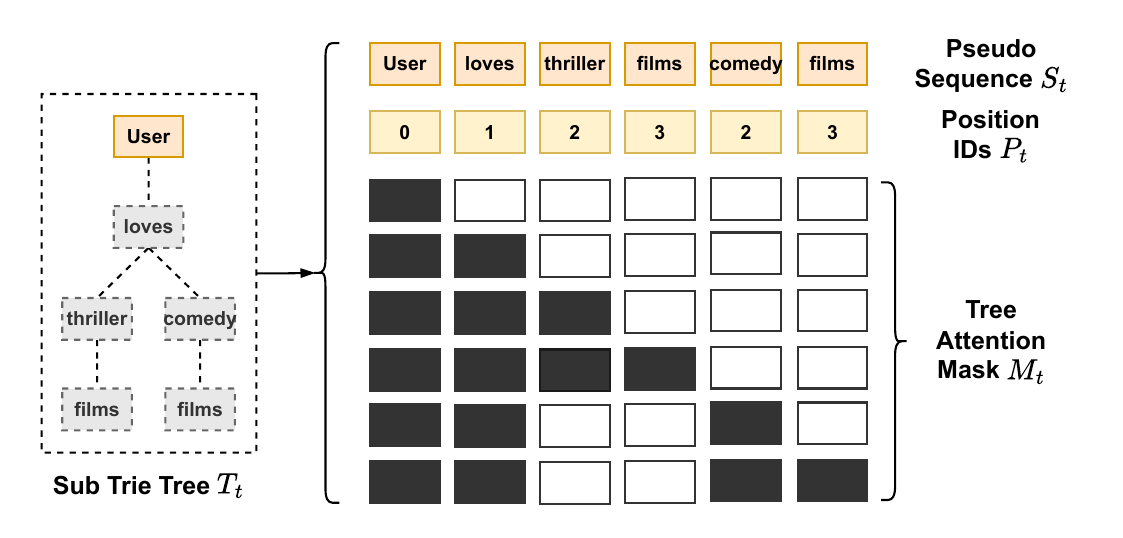}
    \vspace{-13pt}
    \caption{Tree attention.}
    \label{fig:tree_att}
\end{figure}

To reduce the number of decoding steps and increase the possibility of draft tokens being accepted, we aim to validate multiple possible draft sequences from the token tree $T_t$ in a single forward pass of the target LLM. Thus, we utilize the tree attention~\cite{zhao2023lookahead,miao2024specinfer} commonly employed in speculative decoding to validate multiple potential draft sequences in parallel, as illustrated in Figure~\ref{fig:tree_att}. This mechanism constructs a pseudo-sequence $S_t=\{\tilde{x}_{1},\ldots,\tilde{x}_{K}\}$ for token tree $T_t$ with a depth-first search algorithm. Note that the length of $S_t$ may not always reach the maximum length $K$; here, we use $K$ for simplicity. Concurrently, it adjusts the attention mask $M_t$ and position IDs $P_t$ so that each node in the token tree can only see the preceding nodes on the current branch, ensuring that draft sequences from different branches do not interfere with each other.

\subsection{Relaxed Verification}
At this stage, the target LLM will input the original input sequence $\{x_1,\ldots,x_t\}$, the pseudo-sequence $S_t=\{\tilde{x}_{1},\ldots,\tilde{x}_{K}\}$, tree attention mask $M_t$, and position IDs $P_t$ obtained during the drafting phase. It then performs a single forward for parallel validation of all the draft sequences, yielding conditional probability at each position:
\begin{equation}
    q_i = \mathcal{M}(x_{\le t}, \tilde{x}_{< i}, P_t, M_t),\, i=1,\ldots, K+1,
\end{equation}
where $q_i$ denotes the probability distribution of all the tokens in the vocabulary, and $\mathcal{M}$ is the target LLM. In strict/greedy verification, we start from the first position and check if the token $\tilde{x}_i$ at the current position $i$ equals the token with the highest probability in $q_i$ following Eq~\eqref{eq:GR}. If they match, we accept the token $\tilde{x}_i$; otherwise, we reject it. Similarly, if a predecessor node in the token tree $T_t$ is rejected, all of its successor nodes will be skipped. We then proceed to validate the next feasible branch, ultimately accepting the verified branch with the maximum length.

Since we find that recommendation tasks exhibit a high diversity tolerance of LLM-generated knowledge texts in Section~\ref{sec:finding2}, the strict verification could be relaxed to further enhance the generation speed. Therefore, we expand the verification criteria from the highest probability token to the top-$k$ probable tokens, \ie,
\begin{equation}
    \tilde{x}_i \in \text{TopK}(q_i, k),
    \label{eq:topk}
\end{equation}
where the function $\text{TopK}(\cdot)$ selects the tokens with the top-$k$ probabilities in $q_i$. However, our experiments in Section~\ref{sec:ablation} indicate that merely relaxing this constraint can lead to divergent generations, where the text generated in this way is significantly longer than that with autoregressive decoding. This may occur because tokens amongst top-$k$ probabilities, \eg, $e\in\text{TopK}(q_i, k)$, might still have very low actual probabilities $q_i(e)$. Thus, we also impose a probability threshold $p$ to the actual probability and obtain: 
\begin{equation}
\begin{cases} 
\tilde{x}_i \in \text{TopK}(q_i, k), \\
q_i(\tilde{x}_i) > p, \\
\end{cases}
    \label{eq:rv}
\end{equation}
where $q_i(\tilde{x}_i)$ represents the probability of $\tilde{x}_i$ in distribution $q_i$ and the token $\tilde{x}_i$ is accepted only if it meets two conditions in Eq~\eqref{eq:rv}. This relaxed verification enhances the acceptance rate of draft tokens by relaxing the highest probability to the top-$k$ probabilities and effectively prevents divergent generations via the probability threshold $p$, which is validated in Section~\ref{sec:ablation}.

\section{Experiment}

To gain more insights into LASER, we tend to address the following research questions (RQs) in this section. 
\begin{itemize}
    \item \textbf{RQ1:} How does LASER perform in speedup and downstream tasks compared to other speculative decoding approaches?
    \item \textbf{RQ2:} What roles do LASER's two modules, customized retrieval pool and relaxed verification, play in its performance? 
    \item \textbf{RQ3:} How compatible is LASER with different LLMs?
    \item \textbf{RQ4:} What do the draft tokens accepted by LASER look like?
    \item \textbf{RQ5:} What are the performance and costs of deployment?

\end{itemize}

\subsection{Setup}

\subsubsection{Dataset}
Our experiments are conducted on two public datasets, \textbf{MovieLens-10M}\footnote{\url{https://grouplens.org/datasets/movielens/10m/}} and \textbf{Amazon-Books}\footnote{\url{https://cseweb.ucsd.edu/~jmcauley/datasets/amazon_v2/}}. \textbf{MovieLens-10M} (ML-10M for short) contains 10 million movie ratings applied to 10,000 movies by 72,000 users. 
The ratings are converted into binary labels by labeling ratings 4 and 5 as positive and the rest as negative. \textbf{Amazon-Books} is the “Books” category of the Amazon Review Dataset. We filter out the less-interacted users and items, remaining 49,391 users and 78,318 items with 5,002,043 interactions. Ratings of 5 are regarded as positive and the rest as negative. 

The preprocessing of datasets, including knowledge generation and downstream tasks, mainly follows~\cite {xi2023towards}. Additionally, we simulate streaming behaviors and divide the user's historical behaviors into two segments, old and new histories, as mentioned in Section~\ref{sec:finding1}. All items are randomly divided into two equally sized groups: one group as existing items and the other as newly introduced items. To construct retrieval pools, we first employ LLMs to generate old knowledge for users' old histories and existing items with autoregressive decoding. Then, experiments on acceleration and downstream tasks are conducted on new history and new items. 

\subsubsection{Backbone Framework and Baselines}  
As LASER is a model-agnostic decoding strategy, it can accelerate a wide range of recommendation knowledge generation tasks and frameworks. To validate LASER's acceleration performance across different frameworks, we select several typical LLM-based deployable recommendation frameworks, including \textbf{KAR}~\cite{xi2023towards}, \textbf{TRAWL}~\cite{luo2024kellmrec}, \textbf{ONCE}~\cite{liu2024once} and \textbf{RLMRec}~\cite{ren2024representation}.  
These frameworks all extract knowledge from LLMs to enhance traditional RSs. In knowledge extraction, they roughly encompass two major categories of tasks: \textbf{user and item knowledge generation}, despite the specific task instructions may vary, such as user and item profiling or knowledge extraction.

We mainly implement naive retrieval-based speculative decoding (\textbf{ReSD})~\cite{he2023rest} as a baseline because it aligns well with the recommendation knowledge generation scenario without additional model or fine-tuning, and we also verify in Section~\ref{sec:baseline} that it is essentially the SOTA model in this context. This method uses all historical knowledge to construct a prefix tree as a global retrieval pool and employs greedy verification to ensure that generated texts are consistent with autoregressive decoding. It also adopts tree attention to remove the impact of this mechanism on acceleration. Besides, We also compare other representative speculative decoding baselines, such as \textbf{SpecInfer}~\cite{miao2024specinfer}, which uses a small model as the drafter; \textbf{EAGLE}~\cite{li2024eagle}, which employs additional FFN heads for self-drafting; and \textbf{Lookahead}~\cite{fubreak}, which uses Jacobi Iteration for self-drafting.

To validate our model's performance on downstream tasks, we select a crucial task in RSs, \textbf{CTR prediction}, as the downstream task following~\cite{xi2023towards,luo2024kellmrec}, and choose two representative CTR models, \textbf{DIN}~\cite{DIN} and \textbf{DCNv2}~\cite{DCNv2}. The knowledge generated from the four frameworks mentioned above is first encoded by BERT and then adapted to these two models, and we also compare their performance with different speedup strategies, LASER and ReSD.

\subsubsection{Evaluation Metrics } 
For acceleration, we use \textit{Gen. Speed}, which measures the number of tokens generated per second, and \textit{Speedup}, the ratio of the generation speed of the acceleration scheme to that of autoregressive decoding following~\cite{zhao2023lookahead}. During ablation, we adapt \textit{AAL} (average acceptance length), which indicates the average number of draft tokens accepted per decoding step, and \textit{ART} (average retrieval time), representing the average time spent retrieving drafts from retrieval pool for each piece of knowledge. For the downstream task, we employed two commonly used metrics in CTR prediction: \textit{AUC} and \textit{Logloss} (\textit{LL} for short)~\cite{DIN,DCNv2,xi2023towards}.

\subsubsection{Reproducibility.}
All the acceleration experiments on public datasets are conducted on the same NVIDIA RTX 4090 with 24GB memory and 64 CPU cores and all results are averaged over the same set of test samples. Unless specified, LLMs in our experiments refer to Vicuna-7b-v1.3, whose generation speed is 37.4 tokens/s with autoregressive decoding. Our binary router assigns users to a collaborative-based retrieval pool and items to an attribute-based retrieval pool. The collaborative-based retrieval pool adopts K-means clustering with embeddings from LightGCN~\cite{he2020lightgcn}. The number of retrieval pools ranges from 3 to 10 for both approaches. Each group's retrieval pool consists of the previously generated knowledge of the items or users within the group, with the pool size potentially controlled via random sampling. The optimal retrieval pool size may vary across datasets and frameworks, and a grid search within $\{500, 1000, 2000, 3000, 4000, 5000\}$ is performed for optimal size. As for verification, we typically set $k=2$ and $p=0.1$. 

\subsection{Overall Performance (RQ1)}
\subsubsection{Acceleration and Downstream Performance}
These two performances are two key aspects we need to investigate. First, we compare LASER with naive retrieval-based speculative decoding (ReSD) on two tasks (user and item knowledge generation) under four LLM-based recommendation frameworks (KAR, TRAWL, ONCE, and RLMRec). Next, we utilize LASER and ReSD to generate knowledge for all new user histories and items, adapting this knowledge to DIN and DCNv2 according to different frameworks' designs. Note that texts generated by ReSD are utilized as a baseline for downstream task comparison because ReSD employs strict verification, ensuring that its generated results are identical to those of autoregressive generation. Therefore, its performance on downstream tasks is also consistent with autoregressive generation. The above results are presented in Table~\ref{tab:overall}. 

From the acceleration results, we draw the following observations: (i) LASER consistently outperforms ReSD in terms of acceleration across different frameworks and tasks. For instance, in the user knowledge generation task of KAR on Amazon-Books, LASER achieves an acceleration of $4.77\times$ compared to ReSD's $1.71\times$, showcasing an improvement of 178\%. This demonstrates that the two optimizations of LASER significantly enhance speedup performance in generating recommendation knowledge. (ii) The speedup for user knowledge generation is more significant than that for item knowledge generation, with LASER showing greater improvement over ReSD on the user side. LASER achieves an acceleration of 3.86x-4.92x for users, compared to 2.14x-3.28x for items. This may be due to user preferences continuity, resulting in higher similarity between old and new user knowledge. 
\begin{table*}[h]

    
    \caption{Speedup and downstream performance of naive retrieval-based speculative decoding (ReSD) and LASER. }
    \centering
    \scalebox{0.8}{
    \setlength{\tabcolsep}{1.0mm}
    {\begin{tabular}{cc|cccc|cccc|cccc|cccc}
\toprule
\multirow{4}{*}{\begin{tabular}[c]{@{}c@{}}Frame-\\ work\end{tabular}} & \multirow{4}{*}{\begin{tabular}[c]{@{}c@{}}Speedup\\ Method\end{tabular}} & \multicolumn{8}{c|}{ML-10M} & \multicolumn{8}{c}{Amazon-Books} \\
\cmidrule{3-18}
 &  & \multicolumn{4}{c|}{Speedup Performance} & \multicolumn{4}{c|}{Downstream Performance} & \multicolumn{4}{c|}{Speedup Performance} & \multicolumn{4}{c}{Downstream Performance} \\
 \cmidrule{3-18}
 &  & \multicolumn{2}{c}{User Task} & \multicolumn{2}{c|}{Item Task} & \multicolumn{2}{c}{DIN} & \multicolumn{2}{c|}{DCNv2} & \multicolumn{2}{c}{User Task} & \multicolumn{2}{c|}{Item Task} & \multicolumn{2}{c}{DIN} & \multicolumn{2}{c}{DCNv2} \\
 \cmidrule{3-18}
 &  & Gen. Speed & Speedup & Gen. Speed & Speedup & AUC & LL & AUC & LL & Gen. Speed & Speedup & Gen. Speed & Speedup & AUC & LL & AUC & LL \\
 \midrule
base & / & / & / & / & / & 0.8163 & 0.3619 & 0.8115 & 0.3663 & / & / & / & / & 0.8269 & 0.5041 & 0.8241 & 0.5075 \\
\midrule
\multirow{2}{*}{KAR} & ReSD & 94.8 & 2.53$\times$ & 82.8 & 2.21$\times$ & 0.8351 & 0.3469 & 0.8319 & 0.3500 & 64.3 & 1.71$\times$ & 75.2 & 2.01$\times$ & 0.8360 & 0.4962 & 0.8308 & 0.5000 \\
 & LASER & \textbf{171.3} & \textbf{4.58$\times$} & \textbf{107.3} & \textbf{2.87$\times$} & 0.8349 & 0.3474 & 0.8318 & 0.3500 & \textbf{178.9} & \textbf{4.77$\times$} & \textbf{123.0} & \textbf{3.28$\times$} & 0.8358 & 0.4965 & 0.8306 & 0.4996 \\
 \midrule
\multirow{2}{*}{TRAWL} & ReSD & 70.9 & 1.90$\times$ & 81.4 & 2.18$\times$ & 0.8338 & 0.3485 & 0.8314 & 0.3506 & 82.2 & 2.19$\times$ & 76.9 & 2.05$\times$ & 0.8311 & 0.4997 & 0.8301 & 0.5005 \\
 & LASER & \textbf{164.9} & \textbf{4.41$\times$} & \textbf{100.5} & \textbf{2.69$\times$} & 0.8336 & 0.3485 & 0.8314 & 0.3506 & \textbf{144.6} & \textbf{3.86$\times$} & \textbf{120.2} & \textbf{3.21$\times$} & 0.8311 & 0.4998 & 0.8300 & 0.5005 \\
 \midrule
\multirow{2}{*}{ONCE} & ReSD & 68.9 & 1.84$\times$ & 66.3 & 1.77$\times$ & 0.8321 & 0.3511 & 0.8283 & 0.3537 & 71.7 & 1.91$\times$ & 60.1 & 1.60$\times$ & 0.8337 & 0.4952 & 0.8289 & 0.5016 \\
 & LASER & \textbf{154.9} & \textbf{4.14$\times$} & \textbf{80.2} & \textbf{2.14$\times$} & 0.8319 & 0.3511 & 0.8286 & 0.3529 & \textbf{184.5} & \textbf{4.92$\times$} & \textbf{100.1} & \textbf{2.67$\times$} & 0.8332 & 0.4956 & 0.8285 & 0.5017 \\
 \midrule
\multirow{2}{*}{RLMRec} & ReSD & \textit{62.0} & 1.66$\times$ & 85.2 & 2.28$\times$ & 0.8301 & 0.3516 & 0.8281 & 0.3534 & 61.0 & 1.63$\times$ & 56.3 & 1.50$\times$ & 0.8378 & 0.4904 & 0.8325 & 0.4964 \\
 & LASER & \textbf{152.8} & \textbf{4.09$\times$} & \textbf{113.7} & \textbf{3.04$\times$} & 0.8301 & 0.3515 & 0.8282 & 0.3537 & \textbf{150.5} & \textbf{4.01$\times$} & \textbf{116.8} & \textbf{3.11$\times$} & 0.8380 & 0.4903 & 0.8327 & 0.4962
\\ 
\bottomrule
\end{tabular}}
}
\label{tab:overall}
\end{table*}

From the downstream performance, we make the following observations: (i) Knowledge generated by LLMs significantly enhances downstream task performance, with the extent of enhancement varying across frameworks and datasets. For instance, on ML-10M, knowledge from KAR provides a 2.3\% improvement in AUC for DIN. (ii) Across different frameworks, datasets, and backbone CTR models, the performance difference between knowledge generated by LASER and ReSD on downstream tasks is negligible. This indicates that LASER can maintain the performance of downstream tasks while providing significant acceleration of knowledge generation.
\subsubsection{Comparison with Other Speculative Decoding Methods}\label{sec:baseline} 
\begin{figure}[h]
    \centering
    \vspace{-0pt}
    \includegraphics[clip,width=\columnwidth]{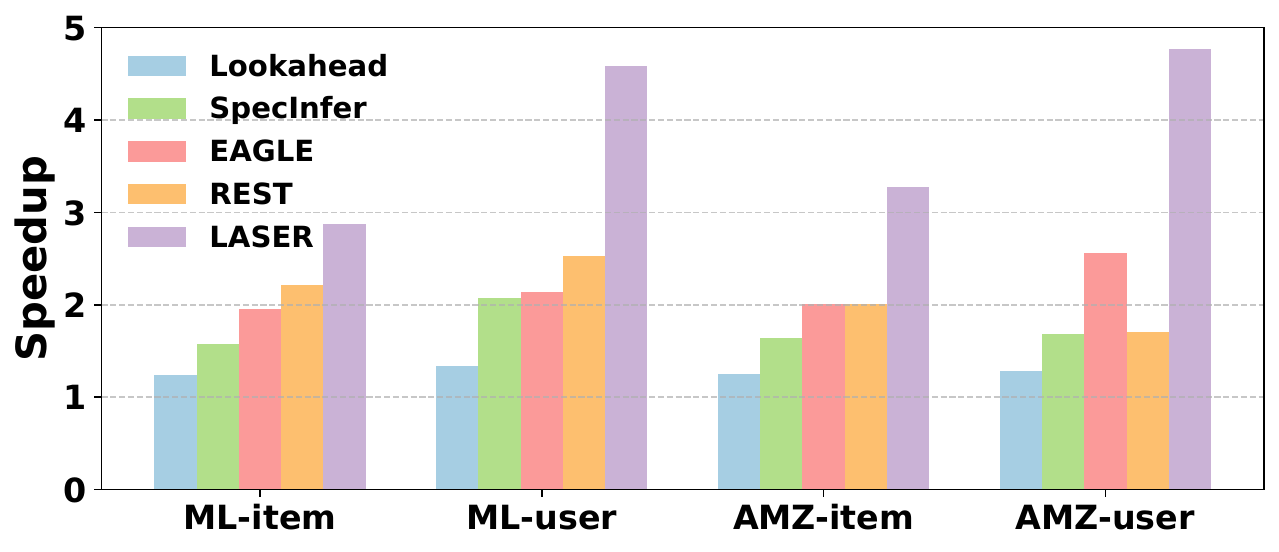}
    \vspace{-20pt}
    \caption{Comparison with speculative decoding methods.}
    \label{fig:baseline}
\end{figure}

To validate the effectiveness of our LASER in acceleration, we compare several representative speculative decoding baselines. These include \textbf{SpecInfer}~\cite{miao2024specinfer}, which uses a small model as the drafter; \textbf{Lookahead}~\cite{fubreak}, which employs Jacobi Iteration for self-drafting; \textbf{REST} (\ie, ReSD)~\cite{he2023rest}, a retrieval-based method. We also include the best-performing model in speculative decoding benchmarks~\cite{xia2024unlocking}, \textbf{EAGLE}~\cite{li2024eagle}, which utilizes additional FFN heads and fine-tuning. With KAR as the backbone framework, we evaluate these baselines alongside LASER on user and item knowledge generation tasks across MovieLens-10M and Amazon-Books datasets (represented as \textbf{ML-item, ML-user, AMZ-item}, and \textbf{AMZ-user} on the x-axis), with the acceleration performance presented in Figure~\ref{fig:baseline}.

The results show that LASER achieves significantly better acceleration than other methods. Furthermore, the retrieval-based method REST often outperforms the SOTA baseline, EAGLE. This indicates that the recommendation knowledge generation scenario is highly suitable for retrieval-based speculative decoding approaches, and LASER's optimizations tailored to recommendation scenarios can further enhance speed and resource efficiency.

\subsection{In-depth Analysis}
\subsubsection{Ablation Study (RQ2)}\label{sec:ablation}
To validate the effectiveness of the two modules we designed in LASER, Customized Retrieval Pool and Relaxed Verification, we conduct ablation and further analysis experiments on them. First, we create several variants for the Customized Retrieval Pool: \textbf{GRP} utilizes all the old knowledge to generate a global retrieval pool, \textbf{CRP} represents our designed Customized Retrieval Pool, and \textbf{RRP} employs random grouping to create retrieval pools with the same size of CRP. These variants are incorporated with greedy verification, whereas \textbf{RV+GRP}, \textbf{LASER}, and \textbf{RV+RRP} are their respective versions enhanced by Relaxed Verification (RV). We examine the performance of these variants and our LASER on user knowledge generation tasks within framework KAR, whose results are presented in Table~\ref{tab:ablation_crp}.
\begin{table}[h]

    \vspace{-0pt}
    \caption{Ablation of LASER.}
    \centering
    \scalebox{0.78}{
    \setlength{\tabcolsep}{1.1mm}
    {\begin{tabular}{c|cccc|cccc}
\toprule
\multirow{2}{*}{Variants} & \multicolumn{4}{c|}{ML-10M} & \multicolumn{4}{c}{Amazon-Books} \\
\cmidrule{2-9}
 & AAL & ART & Gen. Speed & Speedup & AAL & ART & Gen. Speed & Speedup \\
 \midrule
GRP & 6.2 & 1.019 & 94.8 & 2.53$\times$ & 5.64 & 4.120 & 64.3 & 1.71$\times$ \\
RRP & 5.41 & 0.383 & 128.6 & 3.44$\times$ & 5.12 & 0.181 & 128.7 & 3.43$\times$ \\
CRP & 5.51 & 0.121 & 136.3 & 3.64$\times$ & 5.29 & 0.218 & 139.5 & 3.72$\times$ \\
RV+GRP & \textbf{8.32} & 1.976 & 88.7 & 2.37$\times$ & \textbf{7.25} & 7.235 & 57.4 & 1.53$\times$ \\
RV+RRP & 7.41 & 0.215 & 162.2 & 4.34$\times$ & 6.63 & 0.344 & 155.6 & 4.15$\times$ \\
LASER & 7.54 & \textbf{0.064} & \textbf{171.3} & \textbf{4.58$\times$} & 6.76 & \textbf{0.108} & \textbf{178.9} & \textbf{4.77$\times$}
\\ 
\bottomrule
\end{tabular}}
}
\label{tab:ablation_crp}
\end{table}

Firstly, our designed Customized Retrieval Pool (CRP) significantly enhances generation speed, attributed to CRP's ability to reduce Average Retrieval Time (ART) while maintaining a relatively high Average Acceptance Length (AAL),  average number of draft tokens accepted
per decoding step. Compared to the global retrieval pool (GRP), CRP drastically reduces retrieval time from retrieval pools, and it achieves higher AAL than random grouping retrieval pools (RRP) of the same size. This demonstrates that CRP maintains moderate capacity and content similarity. Secondly, Relaxed Verification (RV) boosts token acceptance rates, leading to higher AAL when combined with any retrieval pool. Although the global pool combined with RV (GRP+RV) yields the highest AAL due to its comprehensive content, its large retrieval pool also extends retrieval time, thus hindering overall generation speed. Finally, CRP and RV complement each other; their combination results in reduced retrieval time and higher token acceptance rates. This synergy allows our method, LASER, to achieve a faster generation speed.

\begin{figure}[h]
    \centering
    \includegraphics[clip,width=\columnwidth]{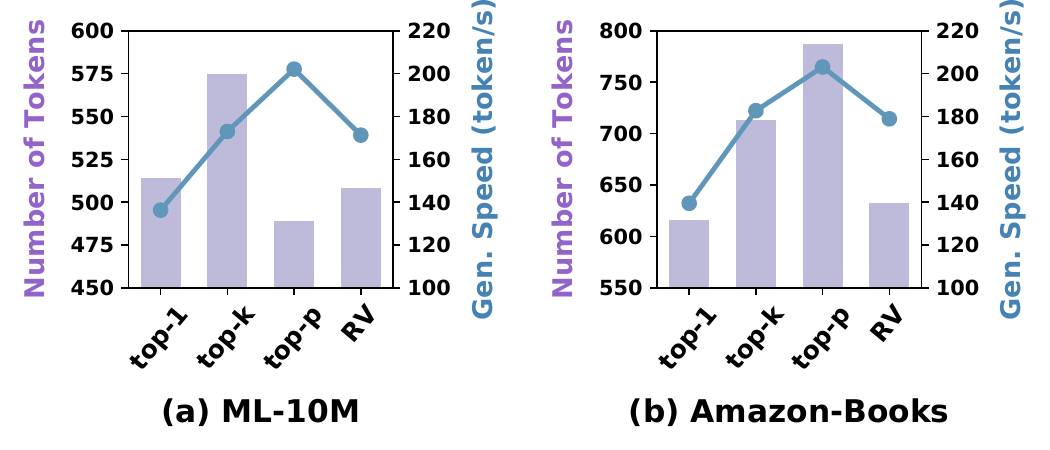}
    \vspace{-15pt}
    \caption{Ablation on Relaxed Verification.}
    \label{fig:abl_vr}
\end{figure}

Next, we delve deeper into relaxed verification by designing several variants: \textbf{top-1} employs greedy verification, \textbf{top-k} satisfies only Eq~\eqref{eq:topk} by accepting tokens with the highest top-$k$ ($k=2$) probabilities, and \textbf{top-p} explores the effect of only meeting the probability threshold, that is accepting tokens with highest probability or absolute probabilities greater than $p$ ($p=0.1$). \textbf{RV} is our designed Relaxed Verification that meets the condition in Eq~\eqref{eq:rv} with $k=2$ and $p=0.1$. We investigate the difference in the number of generated tokens and generation speed on the user knowledge generation task within the framework KAR, as shown in Figure~\ref{fig:abl_vr}.

The results show that compared to the original top-1, both top-k, top-p, and RV significantly enhance generation speed. However, top-k suffers from divergent generation issues, leading to a substantial increase in the number of generated tokens. This ultimately lengthens the total generation time, sometimes surpassing that of top-1, thus failing to achieve practical acceleration. Top-p's performance is inconsistent; it causes divergent generation in Amazon-Books, whereas in ML-10M, the length of the generated text is notably shortened. By combining these approaches, RV achieves a similar length of text generated to that of top-1 while maintaining a generation speed similar to top-k.

\subsubsection{Case Study (RQ4)}\label{sec:case_study}
To investigate which draft tokens LASER actually accepts, we randomly select a user from the ML-10M dataset and employ Vicuna-7b-v1.3 to generate user preference knowledge with LASER. We utilize the prompt of KAR~\cite{xi2023towards} to generate user knowledge, and the results are as follows, where we use \textcolor{red}{\textbf{|}}  to separate the content generated at each decoding step:
\vspace{-5pt}

\begin{tcolorbox}[colback=white, colframe=white, width=0.5\textwidth, title={}]
\textit{Based \textcolor{red}{\textbf{|}}  on the user's movie viewing history and other relevant factors \textcolor{red}{\textbf{|}}  , their preferences can be analyzed as follows: }

\textit{\textcolor{red}{\textbf{|}}  1. Genre: The user seems to enjoy a variety of \textcolor{red}{\textbf{|}}  genres, including action, comedy, drama, and \textcolor{red}{\textbf{|}}  science fiction \textcolor{red}{\textbf{|}}  .}

\textit{2. Directors/Actors: They have a \textcolor{red}{\textbf{|}}  preference for movies directed by and starring famous directors and \textcolor{red}{\textbf{|}}  actors such as Quentin Tarantino \textcolor{red}{\textbf{|}}  ,Kevin Spacey, and Morgan \textcolor{red}{\textbf{|}}  Freeman.}

\textit{3. Time Period/Country: The user \textcolor{red}{\textbf{|}}  enjoys movies from various time periods and countries, including \textcolor{red}{\textbf{|}}  classics from the 1990s and newer releases from \textcolor{red}{\textbf{|}}  the 2000s and 2010s \textcolor{red}{\textbf{|}}  .}

\textit{4. Characters: The user \textcolor{red}{\textbf{|}}  seems to appreciate movies with \textcolor{red}{\textbf{|}}  strong, memorable \textcolor{red}{\textbf{|}}  characters, such as those \textcolor{red}{\textbf{|}}  featuring Ace \textcolor{red}{\textbf{|}}  Ventura, Forrest Gump, and Leon \textcolor{red}{\textbf{|}} .}

\textit{5. Plot/Theme: The user enjoys movies with \textcolor{red}{\textbf{|}}  engaging plots and themes, \textcolor{red}{\textbf{|}}  such as those with \textcolor{red}{\textbf{|}}  elements of mystery, \textcolor{red}{\textbf{|}}  adventure, and drama.}

\textit{6 \textcolor{red}{\textbf{|}}  . Mood/Tone: The user prefers movies \textcolor{red}{\textbf{|}}  with a variety \textcolor{red}{\textbf{|}}  of moods and tones, including comedies, \textcolor{red}{\textbf{|}}  dramas \textcolor{red}{\textbf{|}}  , and thrillers.}

\textit{7. Critical Acclaim/Award: The user seems to appreciate movies that \textcolor{red}{\textbf{|}}  have received critical acclaim and awards, such as \textcolor{red}{\textbf{|}} 12 Monkeys, The Shawshank Redemption, and \textcolor{red}{\textbf{|}}  The Fugitive.} \textcolor{red}{\textbf{|}} 

\textit{8. Production Quality: The user enjoys movies with \textcolor{red}{\textbf{|}}  high production quality, as evidenced by their favorites like \textcolor{red}{\textbf{|}}  Braveheart and The Rock. \textcolor{red}{\textbf{|}}  }

\textit{9 \textcolor{red}{\textbf{|}} . Soundtrack: The user seems to appreciate movies with memorable \textcolor{red}{\textbf{|}}  soundtracks, such as Pulp Fiction \textcolor{red}{\textbf{|}}  and Speed \textcolor{red}{\textbf{|}} .}
\end{tcolorbox}
\vspace{-5pt}

It is evident that, in most cases, LASER can generate multiple tokens within a single decoding step. Some of these received tokens pertain to commonly used phrases, movie titles, and actor/director names, while others involve the recombination of key preferences related to user interests, \eg, genre and theme.

\subsubsection{Compatibility Study (RQ3)}\label{sec:compatibility}
Previous experiments involved the compatibility of LASER across different datasets, LLM-based RS frameworks, and tasks. This section investigates the compatibility of LASER with various backbone LLMs. We select some widely used LLMs, \eg, \textbf{Mistral-7B-instruct-v0.2}~\cite{jiang2023mistral}, \textbf{ChatGLM2-6B}~\cite{glm2024chatglm}, \textbf{Vicuna-7B-v1.5}~\cite{zheng2023judging}, \textbf{Qwen-7B-Chat}~\cite{bai2023qwen}, and \textbf{Qwen-1.8B-Chat}~\cite{bai2023qwen}, and present LASER’s acceleration performance on user/item knowledge generation tasks within the framework KAR in Table~\ref{tab:Compatibility}. Note that the autoregressive generation speeds of different LLMs vary from 30-45 tokens/s; we have omit those speeds due to page limitations. Firstly, across various backbone LLMs, our proposed LASER consistently and significantly outperforms ReSD, demonstrating LASER’s strong compatibility with different backbone LLMs. Secondly, LASER usually exhibits better acceleration on larger LLMs. For instance, on Amazon-Book, LASER achieves accelerations of 5.10x and 4.69x for Qwen-7B-Chat, while for Qwen-1.8B-Chat, the accelerations are 4.77x and 3.98x.

\begin{table}[h]

    \vspace{-0pt}
    \caption{Speedup comparison between naive retrieval-based speculative decoding (ReSD) and LASER with various LLMs. }
    \centering
    \scalebox{0.8}{
    \setlength{\tabcolsep}{1.0mm}
    {\begin{tabular}{ccc|cc|cc}
\toprule

\multirow{2}{*}{Backbone LLM} & \multirow{2}{*}{Side} & \multirow{2}{*}{\begin{tabular}[c]{@{}c@{}}Speedup\\ Method\end{tabular}} & \multicolumn{2}{c|}{ML-10M} & \multicolumn{2}{c}{Amazon-Books} \\
\cmidrule{4-7}
 &  &  & Gen. Speed & Speedup & Gen. Speed & Speedup \\
 \midrule
\multirow{4}{*}{Mistral-7B-Instruct} & \multirow{2}{*}{user} & ReSD & 66.10 & 1.60 $\times$ & 91.4 & 2.21$\times$ \\
 &  & \textbf{LASER} & \textbf{179.00} & \textbf{4.33$\times$} & \textbf{178.9} & \textbf{4.32$\times$} \\
 \cmidrule{2-7}
 & \multirow{2}{*}{item} & ReSD & 80.60 & 1.95$\times$ & 59.4 & 1.43$\times$ \\
 &  & \textbf{LASER} & \textbf{95.30} & \textbf{2.31$\times$} & \textbf{121.5} & \textbf{2.93$\times$} \\
 \midrule
\multirow{4}{*}{ChatGLM2-6B} & \multirow{2}{*}{user} & ReSD & 107.40 & 2.52$\times$ & 89.1 & 2.10$\times$ \\
 &  & \textbf{LASER} & \textbf{194.70} & \textbf{4.57$\times$} & \textbf{165.0} & \textbf{3.88$\times$} \\
  \cmidrule{2-7}
 & \multirow{2}{*}{item} & ReSD & 94.30 & 2.21$\times$ & 71.1 & 1.67$\times$ \\
 &  & \textbf{LASER} & \textbf{106.80} & \textbf{2.51$\times$} & \textbf{117.7} & \textbf{2.77$\times$} \\
 \midrule
\multirow{4}{*}{Vicuna-7B-v1.5} & \multirow{2}{*}{user} & ReSD & 101.10 & 2.67$\times$ & 62.4 & 1.70$\times$ \\
 &  & \textbf{LASER} & \textbf{167.80} & \textbf{4.44$\times$} & \textbf{173.3} & \textbf{4.72$\times$} \\
  \cmidrule{2-7}
 & \multirow{2}{*}{item} & ReSD & 91.10 & 2.41$\times$ & 83.3 & 2.27$\times$ \\
 &  & \textbf{LASER} & \textbf{112.20} & \textbf{2.97$\times$} & \textbf{123.9} & \textbf{3.38$\times$} \\
 \midrule
\multirow{4}{*}{Qwen-7B-Chat} & \multirow{2}{*}{user} & ReSD & 88.10 & 2.94$\times$ & 69.60 & 2.33$\times$ \\
 &  & \textbf{LASER} & \textbf{169.00} & \textbf{5.63$\times$} & \textbf{152.40} & \textbf{5.10$\times$} \\
  \cmidrule{2-7}
 & \multirow{2}{*}{item} & ReSD & 71.90 & 2.40$\times$ & 68.70 & 2.30$\times$ \\
 &  & \textbf{LASER} & \textbf{87.10} & \textbf{2.90$\times$} & \textbf{140.10} & \textbf{4.69$\times$} \\
 \midrule
\multirow{4}{*}{Qwen-1.8B-Chat} & \multirow{2}{*}{user} & ReSD & 78.60 & 1.92$\times$ & 105.90 & 2.58$\times$ \\
 &  & \textbf{LASER} & \textbf{220.00} & \textbf{5.38$\times$} & \textbf{196.20} & \textbf{4.77$\times$} \\
  \cmidrule{2-7}
 & \multirow{2}{*}{item} & ReSD & 112.40 & 2.75$\times$ & 76.20 & 1.85$\times$ \\
 &  & \textbf{LASER} & \textbf{149.20} & \textbf{3.65$\times$} & \textbf{163.60} & \textbf{3.98$\times$}

\\ 
\bottomrule
\end{tabular}}
}
\label{tab:Compatibility}
\vspace{-2pt}
\end{table}

\subsection{Online Deployment (RQ5)}\label{sec:online}
Our experiments are conducted in Huawei's commercial advertising scenario with tens of millions of users and ads. First, LLMs are invoked to analyze ads from diverse aspects, \,  e.g., characteristics, potential target audience, and competitive advantages. Then, generated knowledge is encoded and applied to a downstream conversion rate prediction (CVR) model tailored for this scenario. 

\subsubsection{Speedup and Downstream Performance} We first conduct offline experiments on the industrial dataset from this scenario where ReSD and LASER are integrated into the knowledge generation process. With an in-house developed LLM of 7 billion parameters as the backbone, ReSD achieves a 1.37x speedup, while LASER achieves a \textbf{3.23x speedup}, with a 135.8\% improvement over ReSD. Next, we apply knowledge from ReSD and LASER to the downstream CVR model, as presented in Table~\ref{tab:industrial_downstream}. "\textbf{Base}" indicates no knowledge enhancement, while "\textbf{LASER-Emb}" and "\textbf{ReSD-Emb}" means the CVR model directly utilizes the encoding of LLM-generated knowledge as features. "\textbf{LASER-ID}" and "\textbf{ReSD-ID}" refer to a common optimization approach in the industry, where the encoding of generated knowledge is converted into categorical features (ID) through clustering and then used in CVR models. Table~\ref{tab:industrial_downstream} reveals that demonstrates that LASER and RESD exhibit comparable performance on downstream tasks, indicating LASER's ability to achieve lossless speedup and strong potential for industrial deployment.


\begin{table}[h]

    \caption{Downstream performance on industrial scenarios. }
    \vspace{-6pt}
    \centering
    \scalebox{0.9}{
    \setlength{\tabcolsep}{1.50mm}
    {\begin{tabular}{c|ccccc}
\toprule
Method & Base & ReSD-Emb & LASER-Emb & ReSD-ID & LASER-ID \\
\midrule
AUC & 0.7354 & 0.7396 & 0.7393 & 0.7409 & 0.7405 
\\ 
\bottomrule
\end{tabular}}
}
\label{tab:industrial_downstream}
\end{table}

 In a \textbf{two-week online A/B test} in Huawei's advertising scenario, 10\% users are randomly selected for the experimental group and another 10\% for the control group, both with LLM knowledge augmentation. The only difference is that the experimental group is accelerated by LASER, while the control group employs the original autoregressive decoding. LASER saves about \textbf{67\% of computational resources} per day. The CVR models' performance (such as eCPM) remains consistent, showing no negative impact on downstream tasks. Besides, LASER only requires modifying the decoding strategy of offline LLMs without affecting online service, making it easy to extend to other scenarios.

\subsubsection{Analysis of Additional Overhead} The additional steps of LASER is retrieval pool construction, consisting of the grouping and Trie construction. In the industrial scenario, the time taken by LASER, ReSD, and the original autoregressive decoding (Vanilla) for grouping, Trie construction, and knowledge generation on the same devices is shown in Table~\ref{tab:overhead}.

This scenario employs \textbf{attribute-based retrieval pools}, so its grouping overhead is negligible because it has established groups for items and users. Even without them, the grouping cost is also low for collaborative-based retrieval pools that require clustering. The embeddings for clustering can be sourced from the RS itself without additional training. Since recommendation systems continuously train new models to generate recommendations, we can simply use the embeddings from the most recent model for clustering. 

The cost of Trie construction is also small in relation to the knowledge generation time. Since LASER builds smaller parallel retrieval pools, it is much faster than ReSD, which constructs a global retrieval pool with the entire dataset. Overall, LASER performs significantly better than ReSD in terms of construction time, and Trie construction are much faster than knowledge generation. This suggests that LASER introduces minimal overhead for deployment.

\begin{table}[h]
    \vspace{-5pt}
    \caption{Overhead comparison.}
    \vspace{-5pt}
    \centering
    \scalebox{0.92}{
    \setlength{\tabcolsep}{1.0mm}
    {\begin{tabular}{c|ccc}
    \toprule
Model & Grouping & Trie construction & Knowledge generation \\
\midrule
Vanilla & / & / & 23.17h \\
ReSD & / & 728.93s & 16.92h \\
LASER & <1s & 91.31s & 7.26h \\ 
\bottomrule
\end{tabular}}
}
\label{tab:overhead}
\vspace{-5pt}
\end{table}

\section{Related Work}
\subsection{LLM-based Recommendation}
In recent years, numerous studies have emerged applying LLMs to RSs~\cite{li2023large,zhu2023large,chen2023large,fan2023recommender,wu2023survey,liu2023pre,yu2023self,lin2023can}. Based on how LLMs are utilized, LLM-based recommendations can be categorized into two types. 
One type involves employing LLMs directly as recommenders to generate recommendations. Generally, zero-shot LLMs underperform compared to traditional models in recommendation tasks~\cite{chatrec,liu2023chatgpt,dai2023uncovering,xi2024memocrs,hou2024large,wang2023flip,lin2024clickprompt}. However, LLMs fine-tuned on recommendation data often surpass traditional models~\cite{zheng2024adapting,bao2023tallrec,lin2024rella,zhang2023collm,zhu2024collaborative,zheng2024harnessing,dong2024unsupervised,tan2024idgenrec}, such as TALLREC~\cite{bao2023tallrec} and ReLLa~\cite{lin2024rella}. Despite these advancements, deploying LLMs as recommenders poses significant challenges due to their high inference latency, which is incompatible with the low-latency requirements of RSs.
The other line of work leverages LLMs offline as feature enhancers for traditional RSs~\cite{luo2024kellmrec,liu2024once,lyu2023llm,xi2023towards,ren2024representation,ren2024enhancing,du2024disco,tian2024reland,wang2024can}. Many works~\cite{ren2024representation,xi2023towards,luo2024kellmrec,liu2024once} reasons on user and item knowledge and use well-designed adaptor to adapt the knowledge to the recommendation tasks. This approach avoids LLMs' high online serving latency, making it the mainstream method for integrating LLMs into industrial recommender systems.

Our work focuses on the latter, a more deployable approach. We aim to mitigate the high time and resource consumption when using LLMs offline to generate knowledge for large-scale industrial RSs, specifically by introducing speculative decoding.
\subsection{Speculative Decoding}
The inference latency of LLMs is a significant obstacle to their widespread application. This inefficiency primarily stems from the autoregressive nature of LLMs, where only one token is generated per decoding step. To accelerate LLMs' inference, speculative decoding has been proposed~\cite{stern2018blockwise,xia2023speculative,leviathan2023fast}. This method first efficiently drafts multiple tokens and then utilizes the target LLM to verify parallelly, allowing multiple tokens to be generated in a single decoding step~\cite{xia2024unlocking}.
Current research focuses on two main areas: how to draft and how to verify. The former aims to design effective drafters to produce draft tokens meeting the target LLMs' requirements efficiently. This includes retrieving relevant text from databases~\cite{he2023rest,zhao2023lookahead}, generating text with smaller models from the same series~\cite{leviathan2023fast,chen2023accelerating}, using the target LLM for self-drafting~\cite{cai2024medusa,stern2018blockwise,yang2023predictive,santilli2023accelerating}, and employing knowledge distillation to align the target LLM with the drafter~\cite{zhou2023distillspec,li2024eagle,cai2024medusa,stern2018blockwise}. The latter explores how to verify more draft sequences to improve the token acceptance rate, such as token tree verification~\cite{miao2024specinfer,zhao2023lookahead,li2024eagle,he2023rest,cai2024medusa}.

The above works are primarily focused on accelerating general text generation tasks. We find that retrieval-based speculative decoding is particularly suitable for recommendations, and there is potential for further improvement in the acceleration of recommendations. To this end, we have designed two enhancements to further improve the performance of speculative decoding.

\section{Conclusion}
In this work, we identify the issue of inference efficiency during deploying LLM-based recommendations and introduce speculative decoding to accelerate recommendation knowledge generation. Based on characteristics of speculative decoding in recommendations, we design two key optimizations: Customized Retrieval Pool to reduce retrieval time and Relaxed Verification to increase the number of accepted tokens. Experiments demonstrate that LASER achieves a 3-5x speedup with lossless downstream
performance. LASER can be applied to other domains in information retrieval (IR), \eg, knowledge generation in search. Some techniques from LASER can also be applied beyond IR, such as relaxed verification in cases with high diversity tolerance, \eg, article summarization.


\begin{acks}
    The Shanghai Jiao Tong University team is supported by National Key R\&D Program of China (2022ZD0114804), Shanghai Municipal Science and Technology Major Project (2021SHZDZX0102) and National Natural Science Foundation of China (624B2096, 62322603, 62177033). The work is also sponsored by Huawei Innovation Research Program. We thank MindSpore~\cite{mindspore} for its partial support. The author Yunjia Xi is also supported by Wu Wen Jun Honorary Doctoral Scholarship.
\end{acks}

\bibliographystyle{ACM-Reference-Format}
\balance
\bibliography{sample-base}


\begin{thebibliography}{64}


\ifx \showCODEN    \undefined \def \showCODEN     #1{\unskip}     \fi
\ifx \showISBNx    \undefined \def \showISBNx     #1{\unskip}     \fi
\ifx \showISBNxiii \undefined \def \showISBNxiii  #1{\unskip}     \fi
\ifx \showISSN     \undefined \def \showISSN      #1{\unskip}     \fi
\ifx \showLCCN     \undefined \def \showLCCN      #1{\unskip}     \fi
\ifx \shownote     \undefined \def \shownote      #1{#1}          \fi
\ifx \showarticletitle \undefined \def \showarticletitle #1{#1}   \fi
\ifx \showURL      \undefined \def \showURL       {\relax}        \fi
\providecommand\bibfield[2]{#2}
\providecommand\bibinfo[2]{#2}
\providecommand\natexlab[1]{#1}
\providecommand\showeprint[2][]{arXiv:#2}

\bibitem[min(2020)]%
        {mindspore}
 \bibinfo{year}{2020}\natexlab{}.
\newblock \bibinfo{title}{MindSpore}.
\newblock
\urldef\tempurl%
\url{https://www.mindspore.cn/}
\showURL{%
\tempurl}


\bibitem[Bai et~al\mbox{.}(2023)]%
        {bai2023qwen}
\bibfield{author}{\bibinfo{person}{Jinze Bai}, \bibinfo{person}{Shuai Bai}, \bibinfo{person}{Yunfei Chu}, \bibinfo{person}{Zeyu Cui}, \bibinfo{person}{Kai Dang}, \bibinfo{person}{Xiaodong Deng}, \bibinfo{person}{Yang Fan}, \bibinfo{person}{Wenbin Ge}, \bibinfo{person}{Yu Han}, \bibinfo{person}{Fei Huang}, {et~al\mbox{.}}} \bibinfo{year}{2023}\natexlab{}.
\newblock \showarticletitle{Qwen technical report}.
\newblock \bibinfo{journal}{\emph{arXiv preprint arXiv:2309.16609}} (\bibinfo{year}{2023}).
\newblock


\bibitem[Bao et~al\mbox{.}(2023)]%
        {bao2023tallrec}
\bibfield{author}{\bibinfo{person}{Keqin Bao}, \bibinfo{person}{Jizhi Zhang}, \bibinfo{person}{Yang Zhang}, \bibinfo{person}{Wenjie Wang}, \bibinfo{person}{Fuli Feng}, {and} \bibinfo{person}{Xiangnan He}.} \bibinfo{year}{2023}\natexlab{}.
\newblock \showarticletitle{Tallrec: An effective and efficient tuning framework to align large language model with recommendation}. In \bibinfo{booktitle}{\emph{Proceedings of the 17th ACM Conference on Recommender Systems}}. \bibinfo{pages}{1007--1014}.
\newblock


\bibitem[Brinkmann et~al\mbox{.}(2023)]%
        {brinkmann2023product}
\bibfield{author}{\bibinfo{person}{Alexander Brinkmann}, \bibinfo{person}{Roee Shraga}, {and} \bibinfo{person}{Christian Bizer}.} \bibinfo{year}{2023}\natexlab{}.
\newblock \showarticletitle{Product Attribute Value Extraction using Large Language Models}.
\newblock \bibinfo{journal}{\emph{arXiv preprint arXiv:2310.12537}} (\bibinfo{year}{2023}).
\newblock


\bibitem[Bubeck et~al\mbox{.}(2023)]%
        {bubeck2023sparks}
\bibfield{author}{\bibinfo{person}{S{\'e}bastien Bubeck}, \bibinfo{person}{Varun Chandrasekaran}, \bibinfo{person}{Ronen Eldan}, \bibinfo{person}{Johannes Gehrke}, \bibinfo{person}{Eric Horvitz}, \bibinfo{person}{Ece Kamar}, \bibinfo{person}{Peter Lee}, \bibinfo{person}{Yin~Tat Lee}, \bibinfo{person}{Yuanzhi Li}, \bibinfo{person}{Scott Lundberg}, {et~al\mbox{.}}} \bibinfo{year}{2023}\natexlab{}.
\newblock \showarticletitle{Sparks of artificial general intelligence: Early experiments with gpt-4}.
\newblock \bibinfo{journal}{\emph{arXiv preprint arXiv:2303.12712}} (\bibinfo{year}{2023}).
\newblock


\bibitem[Cai et~al\mbox{.}(2024)]%
        {cai2024medusa}
\bibfield{author}{\bibinfo{person}{Tianle Cai}, \bibinfo{person}{Yuhong Li}, \bibinfo{person}{Zhengyang Geng}, \bibinfo{person}{Hongwu Peng}, \bibinfo{person}{Jason~D Lee}, \bibinfo{person}{Deming Chen}, {and} \bibinfo{person}{Tri Dao}.} \bibinfo{year}{2024}\natexlab{}.
\newblock \showarticletitle{Medusa: Simple llm inference acceleration framework with multiple decoding heads}.
\newblock \bibinfo{journal}{\emph{arXiv preprint arXiv:2401.10774}} (\bibinfo{year}{2024}).
\newblock


\bibitem[Chen et~al\mbox{.}(2023a)]%
        {chen2023accelerating}
\bibfield{author}{\bibinfo{person}{Charlie Chen}, \bibinfo{person}{Sebastian Borgeaud}, \bibinfo{person}{Geoffrey Irving}, \bibinfo{person}{Jean-Baptiste Lespiau}, \bibinfo{person}{Laurent Sifre}, {and} \bibinfo{person}{John Jumper}.} \bibinfo{year}{2023}\natexlab{a}.
\newblock \showarticletitle{Accelerating large language model decoding with speculative sampling}.
\newblock \bibinfo{journal}{\emph{arXiv preprint arXiv:2302.01318}} (\bibinfo{year}{2023}).
\newblock


\bibitem[Chen et~al\mbox{.}(2023b)]%
        {chen2023large}
\bibfield{author}{\bibinfo{person}{Jin Chen}, \bibinfo{person}{Zheng Liu}, \bibinfo{person}{Xu Huang}, \bibinfo{person}{Chenwang Wu}, \bibinfo{person}{Qi Liu}, \bibinfo{person}{Gangwei Jiang}, \bibinfo{person}{Yuanhao Pu}, \bibinfo{person}{Yuxuan Lei}, \bibinfo{person}{Xiaolong Chen}, \bibinfo{person}{Xingmei Wang}, {et~al\mbox{.}}} \bibinfo{year}{2023}\natexlab{b}.
\newblock \showarticletitle{When large language models meet personalization: Perspectives of challenges and opportunities}.
\newblock \bibinfo{journal}{\emph{arXiv preprint arXiv:2307.16376}} (\bibinfo{year}{2023}).
\newblock


\bibitem[Dai et~al\mbox{.}(2023)]%
        {dai2023uncovering}
\bibfield{author}{\bibinfo{person}{Sunhao Dai}, \bibinfo{person}{Ninglu Shao}, \bibinfo{person}{Haiyuan Zhao}, \bibinfo{person}{Weijie Yu}, \bibinfo{person}{Zihua Si}, \bibinfo{person}{Chen Xu}, \bibinfo{person}{Zhongxiang Sun}, \bibinfo{person}{Xiao Zhang}, {and} \bibinfo{person}{Jun Xu}.} \bibinfo{year}{2023}\natexlab{}.
\newblock \showarticletitle{Uncovering ChatGPT's Capabilities in Recommender Systems}.
\newblock \bibinfo{journal}{\emph{arXiv preprint arXiv:2305.02182}} (\bibinfo{year}{2023}).
\newblock


\bibitem[De~La~Briandais(1959)]%
        {de1959file}
\bibfield{author}{\bibinfo{person}{Rene De~La~Briandais}.} \bibinfo{year}{1959}\natexlab{}.
\newblock \showarticletitle{File searching using variable length keys}. In \bibinfo{booktitle}{\emph{Papers presented at the the March 3-5, 1959, western joint computer conference}}. \bibinfo{pages}{295--298}.
\newblock


\bibitem[Dong et~al\mbox{.}(2024)]%
        {dong2024unsupervised}
\bibfield{author}{\bibinfo{person}{Qian Dong}, \bibinfo{person}{Yiding Liu}, \bibinfo{person}{Qingyao Ai}, \bibinfo{person}{Zhijing Wu}, \bibinfo{person}{Haitao Li}, \bibinfo{person}{Yiqun Liu}, \bibinfo{person}{Shuaiqiang Wang}, \bibinfo{person}{Dawei Yin}, {and} \bibinfo{person}{Shaoping Ma}.} \bibinfo{year}{2024}\natexlab{}.
\newblock \showarticletitle{Unsupervised large language model alignment for information retrieval via contrastive feedback}. In \bibinfo{booktitle}{\emph{Proceedings of the 47th International ACM SIGIR Conference on Research and Development in Information Retrieval}}. \bibinfo{pages}{48--58}.
\newblock


\bibitem[Du et~al\mbox{.}(2024)]%
        {du2024disco}
\bibfield{author}{\bibinfo{person}{Kounianhua Du}, \bibinfo{person}{Jizheng Chen}, \bibinfo{person}{Jianghao Lin}, \bibinfo{person}{Yunjia Xi}, \bibinfo{person}{Hangyu Wang}, \bibinfo{person}{Xinyi Dai}, \bibinfo{person}{Bo Chen}, \bibinfo{person}{Ruiming Tang}, {and} \bibinfo{person}{Weinan Zhang}.} \bibinfo{year}{2024}\natexlab{}.
\newblock \showarticletitle{DisCo: Towards Harmonious Disentanglement and Collaboration between Tabular and Semantic Space for Recommendation}.
\newblock \bibinfo{journal}{\emph{arXiv preprint arXiv:2406.00011}} (\bibinfo{year}{2024}).
\newblock


\bibitem[Fan et~al\mbox{.}(2023)]%
        {fan2023recommender}
\bibfield{author}{\bibinfo{person}{Wenqi Fan}, \bibinfo{person}{Zihuai Zhao}, \bibinfo{person}{Jiatong Li}, \bibinfo{person}{Yunqing Liu}, \bibinfo{person}{Xiaowei Mei}, \bibinfo{person}{Yiqi Wang}, \bibinfo{person}{Jiliang Tang}, {and} \bibinfo{person}{Qing Li}.} \bibinfo{year}{2023}\natexlab{}.
\newblock \showarticletitle{Recommender systems in the era of large language models (llms)}.
\newblock \bibinfo{journal}{\emph{arXiv preprint arXiv:2307.02046}} (\bibinfo{year}{2023}).
\newblock


\bibitem[Fang et~al\mbox{.}(2024)]%
        {fang2024llm}
\bibfield{author}{\bibinfo{person}{Chenhao Fang}, \bibinfo{person}{Xiaohan Li}, \bibinfo{person}{Zezhong Fan}, \bibinfo{person}{Jianpeng Xu}, \bibinfo{person}{Kaushiki Nag}, \bibinfo{person}{Evren Korpeoglu}, \bibinfo{person}{Sushant Kumar}, {and} \bibinfo{person}{Kannan Achan}.} \bibinfo{year}{2024}\natexlab{}.
\newblock \showarticletitle{LLM-Ensemble: Optimal Large Language Model Ensemble Method for E-commerce Product Attribute Value Extraction}. In \bibinfo{booktitle}{\emph{Proceedings of the 47th International ACM SIGIR Conference on Research and Development in Information Retrieval}}. \bibinfo{pages}{2910--2914}.
\newblock


\bibitem[Fu et~al\mbox{.}(2024)]%
        {fubreak}
\bibfield{author}{\bibinfo{person}{Yichao Fu}, \bibinfo{person}{Peter Bailis}, \bibinfo{person}{Ion Stoica}, {and} \bibinfo{person}{Hao Zhang}.} \bibinfo{year}{2024}\natexlab{}.
\newblock \showarticletitle{Break the Sequential Dependency of LLM Inference Using Lookahead Decoding}. In \bibinfo{booktitle}{\emph{Forty-first International Conference on Machine Learning}}.
\newblock


\bibitem[Gao et~al\mbox{.}(2023)]%
        {chatrec}
\bibfield{author}{\bibinfo{person}{Yunfan Gao}, \bibinfo{person}{Tao Sheng}, \bibinfo{person}{Youlin Xiang}, \bibinfo{person}{Yun Xiong}, \bibinfo{person}{Haofen Wang}, {and} \bibinfo{person}{Jiawei Zhang}.} \bibinfo{year}{2023}\natexlab{}.
\newblock \showarticletitle{Chat-REC: Towards Interactive and Explainable LLMs-Augmented Recommender System}.
\newblock \bibinfo{journal}{\emph{arXiv preprint arXiv:2303.14524}} (\bibinfo{year}{2023}).
\newblock


\bibitem[GLM et~al\mbox{.}(2024)]%
        {glm2024chatglm}
\bibfield{author}{\bibinfo{person}{Team GLM}, \bibinfo{person}{Aohan Zeng}, \bibinfo{person}{Bin Xu}, \bibinfo{person}{Bowen Wang}, \bibinfo{person}{Chenhui Zhang}, \bibinfo{person}{Da Yin}, \bibinfo{person}{Dan Zhang}, \bibinfo{person}{Diego Rojas}, \bibinfo{person}{Guanyu Feng}, \bibinfo{person}{Hanlin Zhao}, {et~al\mbox{.}}} \bibinfo{year}{2024}\natexlab{}.
\newblock \showarticletitle{Chatglm: A family of large language models from glm-130b to glm-4 all tools}.
\newblock \bibinfo{journal}{\emph{arXiv preprint arXiv:2406.12793}} (\bibinfo{year}{2024}).
\newblock


\bibitem[He et~al\mbox{.}(2020)]%
        {he2020lightgcn}
\bibfield{author}{\bibinfo{person}{Xiangnan He}, \bibinfo{person}{Kuan Deng}, \bibinfo{person}{Xiang Wang}, \bibinfo{person}{Yan Li}, \bibinfo{person}{Yongdong Zhang}, {and} \bibinfo{person}{Meng Wang}.} \bibinfo{year}{2020}\natexlab{}.
\newblock \showarticletitle{Lightgcn: Simplifying and powering graph convolution network for recommendation}. In \bibinfo{booktitle}{\emph{Proceedings of the 43rd International ACM SIGIR conference on research and development in Information Retrieval}}. \bibinfo{pages}{639--648}.
\newblock


\bibitem[He et~al\mbox{.}(2023)]%
        {he2023rest}
\bibfield{author}{\bibinfo{person}{Zhenyu He}, \bibinfo{person}{Zexuan Zhong}, \bibinfo{person}{Tianle Cai}, \bibinfo{person}{Jason~D Lee}, {and} \bibinfo{person}{Di He}.} \bibinfo{year}{2023}\natexlab{}.
\newblock \showarticletitle{Rest: Retrieval-based speculative decoding}.
\newblock \bibinfo{journal}{\emph{arXiv preprint arXiv:2311.08252}} (\bibinfo{year}{2023}).
\newblock


\bibitem[Hou et~al\mbox{.}(2024)]%
        {hou2024large}
\bibfield{author}{\bibinfo{person}{Yupeng Hou}, \bibinfo{person}{Junjie Zhang}, \bibinfo{person}{Zihan Lin}, \bibinfo{person}{Hongyu Lu}, \bibinfo{person}{Ruobing Xie}, \bibinfo{person}{Julian McAuley}, {and} \bibinfo{person}{Wayne~Xin Zhao}.} \bibinfo{year}{2024}\natexlab{}.
\newblock \showarticletitle{Large language models are zero-shot rankers for recommender systems}. In \bibinfo{booktitle}{\emph{European Conference on Information Retrieval}}. Springer, \bibinfo{pages}{364--381}.
\newblock


\bibitem[Jiang et~al\mbox{.}(2023)]%
        {jiang2023mistral}
\bibfield{author}{\bibinfo{person}{Albert~Q Jiang}, \bibinfo{person}{Alexandre Sablayrolles}, \bibinfo{person}{Arthur Mensch}, \bibinfo{person}{Chris Bamford}, \bibinfo{person}{Devendra~Singh Chaplot}, \bibinfo{person}{Diego de~las Casas}, \bibinfo{person}{Florian Bressand}, \bibinfo{person}{Gianna Lengyel}, \bibinfo{person}{Guillaume Lample}, \bibinfo{person}{Lucile Saulnier}, {et~al\mbox{.}}} \bibinfo{year}{2023}\natexlab{}.
\newblock \showarticletitle{Mistral 7B}.
\newblock \bibinfo{journal}{\emph{arXiv preprint arXiv:2310.06825}} (\bibinfo{year}{2023}).
\newblock


\bibitem[Krishna and Murty(1999)]%
        {krishna1999genetic}
\bibfield{author}{\bibinfo{person}{K Krishna} {and} \bibinfo{person}{M~Narasimha Murty}.} \bibinfo{year}{1999}\natexlab{}.
\newblock \showarticletitle{Genetic K-means algorithm}.
\newblock \bibinfo{journal}{\emph{IEEE Transactions on Systems, Man, and Cybernetics, Part B (Cybernetics)}} \bibinfo{volume}{29}, \bibinfo{number}{3} (\bibinfo{year}{1999}), \bibinfo{pages}{433--439}.
\newblock


\bibitem[Leviathan et~al\mbox{.}(2023)]%
        {leviathan2023fast}
\bibfield{author}{\bibinfo{person}{Yaniv Leviathan}, \bibinfo{person}{Matan Kalman}, {and} \bibinfo{person}{Yossi Matias}.} \bibinfo{year}{2023}\natexlab{}.
\newblock \showarticletitle{Fast inference from transformers via speculative decoding}. In \bibinfo{booktitle}{\emph{International Conference on Machine Learning}}. PMLR, \bibinfo{pages}{19274--19286}.
\newblock


\bibitem[Li et~al\mbox{.}(2023a)]%
        {li2023taggpt}
\bibfield{author}{\bibinfo{person}{Chen Li}, \bibinfo{person}{Yixiao Ge}, \bibinfo{person}{Jiayong Mao}, \bibinfo{person}{Dian Li}, {and} \bibinfo{person}{Ying Shan}.} \bibinfo{year}{2023}\natexlab{a}.
\newblock \showarticletitle{TagGPT: Large Language Models are Zero-shot Multimodal Taggers}.
\newblock \bibinfo{journal}{\emph{arXiv preprint arXiv:2304.03022}} (\bibinfo{year}{2023}).
\newblock


\bibitem[Li et~al\mbox{.}(2023b)]%
        {li2023large}
\bibfield{author}{\bibinfo{person}{Lei Li}, \bibinfo{person}{Yongfeng Zhang}, \bibinfo{person}{Dugang Liu}, {and} \bibinfo{person}{Li Chen}.} \bibinfo{year}{2023}\natexlab{b}.
\newblock \showarticletitle{Large Language Models for Generative Recommendation: A Survey and Visionary Discussions}.
\newblock \bibinfo{journal}{\emph{arXiv preprint arXiv:2309.01157}} (\bibinfo{year}{2023}).
\newblock


\bibitem[Li et~al\mbox{.}(2024)]%
        {li2024eagle}
\bibfield{author}{\bibinfo{person}{Yuhui Li}, \bibinfo{person}{Fangyun Wei}, \bibinfo{person}{Chao Zhang}, {and} \bibinfo{person}{Hongyang Zhang}.} \bibinfo{year}{2024}\natexlab{}.
\newblock \showarticletitle{Eagle: Speculative sampling requires rethinking feature uncertainty}.
\newblock \bibinfo{journal}{\emph{arXiv preprint arXiv:2401.15077}} (\bibinfo{year}{2024}).
\newblock


\bibitem[Lin et~al\mbox{.}(2024a)]%
        {lin2024clickprompt}
\bibfield{author}{\bibinfo{person}{Jianghao Lin}, \bibinfo{person}{Bo Chen}, \bibinfo{person}{Hangyu Wang}, \bibinfo{person}{Yunjia Xi}, \bibinfo{person}{Yanru Qu}, \bibinfo{person}{Xinyi Dai}, \bibinfo{person}{Kangning Zhang}, \bibinfo{person}{Ruiming Tang}, \bibinfo{person}{Yong Yu}, {and} \bibinfo{person}{Weinan Zhang}.} \bibinfo{year}{2024}\natexlab{a}.
\newblock \showarticletitle{ClickPrompt: CTR Models are Strong Prompt Generators for Adapting Language Models to CTR Prediction}. In \bibinfo{booktitle}{\emph{Proceedings of the ACM on Web Conference 2024}}. \bibinfo{pages}{3319--3330}.
\newblock


\bibitem[Lin et~al\mbox{.}(2023)]%
        {lin2023can}
\bibfield{author}{\bibinfo{person}{Jianghao Lin}, \bibinfo{person}{Xinyi Dai}, \bibinfo{person}{Yunjia Xi}, \bibinfo{person}{Weiwen Liu}, \bibinfo{person}{Bo Chen}, \bibinfo{person}{Xiangyang Li}, \bibinfo{person}{Chenxu Zhu}, \bibinfo{person}{Huifeng Guo}, \bibinfo{person}{Yong Yu}, \bibinfo{person}{Ruiming Tang}, {et~al\mbox{.}}} \bibinfo{year}{2023}\natexlab{}.
\newblock \showarticletitle{How Can Recommender Systems Benefit from Large Language Models: A Survey}.
\newblock \bibinfo{journal}{\emph{arXiv preprint arXiv:2306.05817}} (\bibinfo{year}{2023}).
\newblock


\bibitem[Lin et~al\mbox{.}(2024b)]%
        {lin2024rella}
\bibfield{author}{\bibinfo{person}{Jianghao Lin}, \bibinfo{person}{Rong Shan}, \bibinfo{person}{Chenxu Zhu}, \bibinfo{person}{Kounianhua Du}, \bibinfo{person}{Bo Chen}, \bibinfo{person}{Shigang Quan}, \bibinfo{person}{Ruiming Tang}, \bibinfo{person}{Yong Yu}, {and} \bibinfo{person}{Weinan Zhang}.} \bibinfo{year}{2024}\natexlab{b}.
\newblock \showarticletitle{Rella: Retrieval-enhanced large language models for lifelong sequential behavior comprehension in recommendation}. In \bibinfo{booktitle}{\emph{Proceedings of the ACM on Web Conference 2024}}. \bibinfo{pages}{3497--3508}.
\newblock


\bibitem[Liu et~al\mbox{.}(2023a)]%
        {liu2023chatgpt}
\bibfield{author}{\bibinfo{person}{Junling Liu}, \bibinfo{person}{Chao Liu}, \bibinfo{person}{Renjie Lv}, \bibinfo{person}{Kang Zhou}, {and} \bibinfo{person}{Yan Zhang}.} \bibinfo{year}{2023}\natexlab{a}.
\newblock \showarticletitle{Is ChatGPT a Good Recommender? A Preliminary Study}.
\newblock \bibinfo{journal}{\emph{arXiv preprint arXiv:2304.10149}} (\bibinfo{year}{2023}).
\newblock


\bibitem[Liu et~al\mbox{.}(2023b)]%
        {liu2023pre}
\bibfield{author}{\bibinfo{person}{Peng Liu}, \bibinfo{person}{Lemei Zhang}, {and} \bibinfo{person}{Jon~Atle Gulla}.} \bibinfo{year}{2023}\natexlab{b}.
\newblock \showarticletitle{Pre-train, prompt and recommendation: A comprehensive survey of language modelling paradigm adaptations in recommender systems}.
\newblock \bibinfo{journal}{\emph{arXiv preprint arXiv:2302.03735}} (\bibinfo{year}{2023}).
\newblock


\bibitem[Liu et~al\mbox{.}(2024a)]%
        {liu2024once}
\bibfield{author}{\bibinfo{person}{Qijiong Liu}, \bibinfo{person}{Nuo Chen}, \bibinfo{person}{Tetsuya Sakai}, {and} \bibinfo{person}{Xiao-Ming Wu}.} \bibinfo{year}{2024}\natexlab{a}.
\newblock \showarticletitle{Once: Boosting content-based recommendation with both open-and closed-source large language models}. In \bibinfo{booktitle}{\emph{Proceedings of the 17th ACM International Conference on Web Search and Data Mining}}. \bibinfo{pages}{452--461}.
\newblock


\bibitem[Liu et~al\mbox{.}(2024b)]%
        {liu2024modeling}
\bibfield{author}{\bibinfo{person}{Zhenghao Liu}, \bibinfo{person}{Zulong Chen}, \bibinfo{person}{Moufeng Zhang}, \bibinfo{person}{Shaoyang Duan}, \bibinfo{person}{Hong Wen}, \bibinfo{person}{Liangyue Li}, \bibinfo{person}{Nan Li}, \bibinfo{person}{Yu Gu}, {and} \bibinfo{person}{Ge Yu}.} \bibinfo{year}{2024}\natexlab{b}.
\newblock \showarticletitle{Modeling User Viewing Flow using Large Language Models for Article Recommendation}. In \bibinfo{booktitle}{\emph{Companion Proceedings of the ACM on Web Conference 2024}}. \bibinfo{pages}{83--92}.
\newblock


\bibitem[Luo et~al\mbox{.}(2024)]%
        {luo2024kellmrec}
\bibfield{author}{\bibinfo{person}{Weiqing Luo}, \bibinfo{person}{Chonggang Song}, \bibinfo{person}{Lingling Yi}, {and} \bibinfo{person}{Gong Cheng}.} \bibinfo{year}{2024}\natexlab{}.
\newblock \showarticletitle{KELLMRec: Knowledge-Enhanced Large Language Models for Recommendation}.
\newblock \bibinfo{journal}{\emph{arXiv preprint arXiv:2403.06642}} (\bibinfo{year}{2024}).
\newblock


\bibitem[Lyu et~al\mbox{.}(2023)]%
        {lyu2023llm}
\bibfield{author}{\bibinfo{person}{Hanjia Lyu}, \bibinfo{person}{Song Jiang}, \bibinfo{person}{Hanqing Zeng}, \bibinfo{person}{Yinglong Xia}, {and} \bibinfo{person}{Jiebo Luo}.} \bibinfo{year}{2023}\natexlab{}.
\newblock \showarticletitle{Llm-rec: Personalized recommendation via prompting large language models}.
\newblock \bibinfo{journal}{\emph{arXiv preprint arXiv:2307.15780}} (\bibinfo{year}{2023}).
\newblock


\bibitem[Miao et~al\mbox{.}(2024)]%
        {miao2024specinfer}
\bibfield{author}{\bibinfo{person}{Xupeng Miao}, \bibinfo{person}{Gabriele Oliaro}, \bibinfo{person}{Zhihao Zhang}, \bibinfo{person}{Xinhao Cheng}, \bibinfo{person}{Zeyu Wang}, \bibinfo{person}{Zhengxin Zhang}, \bibinfo{person}{Rae Ying~Yee Wong}, \bibinfo{person}{Alan Zhu}, \bibinfo{person}{Lijie Yang}, \bibinfo{person}{Xiaoxiang Shi}, {et~al\mbox{.}}} \bibinfo{year}{2024}\natexlab{}.
\newblock \showarticletitle{Specinfer: Accelerating large language model serving with tree-based speculative inference and verification}. In \bibinfo{booktitle}{\emph{Proceedings of the 29th ACM International Conference on Architectural Support for Programming Languages and Operating Systems, Volume 3}}. \bibinfo{pages}{932--949}.
\newblock


\bibitem[OpenAI(2023)]%
        {gpt4}
\bibfield{author}{\bibinfo{person}{OpenAI}.} \bibinfo{year}{2023}\natexlab{}.
\newblock \showarticletitle{{GPT-4} Technical Report}.
\newblock \bibinfo{journal}{\emph{CoRR}}  \bibinfo{volume}{abs/2303.08774} (\bibinfo{year}{2023}).
\newblock
\urldef\tempurl%
\url{https://doi.org/10.48550/arXiv.2303.08774}
\showURL{%
\tempurl}


\bibitem[Ren et~al\mbox{.}(2024b)]%
        {ren2024representation}
\bibfield{author}{\bibinfo{person}{Xubin Ren}, \bibinfo{person}{Wei Wei}, \bibinfo{person}{Lianghao Xia}, \bibinfo{person}{Lixin Su}, \bibinfo{person}{Suqi Cheng}, \bibinfo{person}{Junfeng Wang}, \bibinfo{person}{Dawei Yin}, {and} \bibinfo{person}{Chao Huang}.} \bibinfo{year}{2024}\natexlab{b}.
\newblock \showarticletitle{Representation learning with large language models for recommendation}. In \bibinfo{booktitle}{\emph{Proceedings of the ACM on Web Conference 2024}}. \bibinfo{pages}{3464--3475}.
\newblock


\bibitem[Ren et~al\mbox{.}(2024a)]%
        {ren2024enhancing}
\bibfield{author}{\bibinfo{person}{Yankun Ren}, \bibinfo{person}{Zhongde Chen}, \bibinfo{person}{Xinxing Yang}, \bibinfo{person}{Longfei Li}, \bibinfo{person}{Cong Jiang}, \bibinfo{person}{Lei Cheng}, \bibinfo{person}{Bo Zhang}, \bibinfo{person}{Linjian Mo}, {and} \bibinfo{person}{Jun Zhou}.} \bibinfo{year}{2024}\natexlab{a}.
\newblock \showarticletitle{Enhancing Sequential Recommenders with Augmented Knowledge from Aligned Large Language Models}. In \bibinfo{booktitle}{\emph{Proceedings of the 47th International ACM SIGIR Conference on Research and Development in Information Retrieval}}. \bibinfo{pages}{345--354}.
\newblock


\bibitem[Santilli et~al\mbox{.}(2023)]%
        {santilli2023accelerating}
\bibfield{author}{\bibinfo{person}{Andrea Santilli}, \bibinfo{person}{Silvio Severino}, \bibinfo{person}{Emilian Postolache}, \bibinfo{person}{Valentino Maiorca}, \bibinfo{person}{Michele Mancusi}, \bibinfo{person}{Riccardo Marin}, {and} \bibinfo{person}{Emanuele Rodol{\`a}}.} \bibinfo{year}{2023}\natexlab{}.
\newblock \showarticletitle{Accelerating transformer inference for translation via parallel decoding}.
\newblock \bibinfo{journal}{\emph{arXiv preprint arXiv:2305.10427}} (\bibinfo{year}{2023}).
\newblock


\bibitem[Stern et~al\mbox{.}(2018)]%
        {stern2018blockwise}
\bibfield{author}{\bibinfo{person}{Mitchell Stern}, \bibinfo{person}{Noam Shazeer}, {and} \bibinfo{person}{Jakob Uszkoreit}.} \bibinfo{year}{2018}\natexlab{}.
\newblock \showarticletitle{Blockwise parallel decoding for deep autoregressive models}.
\newblock \bibinfo{journal}{\emph{Advances in Neural Information Processing Systems}}  \bibinfo{volume}{31} (\bibinfo{year}{2018}).
\newblock


\bibitem[Tan et~al\mbox{.}(2024)]%
        {tan2024idgenrec}
\bibfield{author}{\bibinfo{person}{Juntao Tan}, \bibinfo{person}{Shuyuan Xu}, \bibinfo{person}{Wenyue Hua}, \bibinfo{person}{Yingqiang Ge}, \bibinfo{person}{Zelong Li}, {and} \bibinfo{person}{Yongfeng Zhang}.} \bibinfo{year}{2024}\natexlab{}.
\newblock \showarticletitle{Idgenrec: Llm-recsys alignment with textual id learning}. In \bibinfo{booktitle}{\emph{Proceedings of the 47th International ACM SIGIR Conference on Research and Development in Information Retrieval}}. \bibinfo{pages}{355--364}.
\newblock


\bibitem[Tian et~al\mbox{.}(2024)]%
        {tian2024reland}
\bibfield{author}{\bibinfo{person}{Changxin Tian}, \bibinfo{person}{Binbin Hu}, \bibinfo{person}{Chunjing Gan}, \bibinfo{person}{Haoyu Chen}, \bibinfo{person}{Zhuo Zhang}, \bibinfo{person}{Li Yu}, \bibinfo{person}{Ziqi Liu}, \bibinfo{person}{Zhiqiang Zhang}, \bibinfo{person}{Jun Zhou}, {and} \bibinfo{person}{Jiawei Chen}.} \bibinfo{year}{2024}\natexlab{}.
\newblock \showarticletitle{ReLand: Integrating Large Language Models' Insights into Industrial Recommenders via a Controllable Reasoning Pool}. In \bibinfo{booktitle}{\emph{Proceedings of the 18th ACM Conference on Recommender Systems}}. \bibinfo{pages}{63--73}.
\newblock


\bibitem[Wang et~al\mbox{.}(2023)]%
        {wang2023flip}
\bibfield{author}{\bibinfo{person}{Hangyu Wang}, \bibinfo{person}{Jianghao Lin}, \bibinfo{person}{Xiangyang Li}, \bibinfo{person}{Bo Chen}, \bibinfo{person}{Chenxu Zhu}, \bibinfo{person}{Ruiming Tang}, \bibinfo{person}{Weinan Zhang}, {and} \bibinfo{person}{Yong Yu}.} \bibinfo{year}{2023}\natexlab{}.
\newblock \showarticletitle{FLIP: Towards Fine-grained Alignment between ID-based Models and Pretrained Language Models for CTR Prediction}.
\newblock \bibinfo{journal}{\emph{arXiv e-prints}} (\bibinfo{year}{2023}), \bibinfo{pages}{arXiv--2310}.
\newblock


\bibitem[Wang et~al\mbox{.}(2021)]%
        {DCNv2}
\bibfield{author}{\bibinfo{person}{Ruoxi Wang}, \bibinfo{person}{Rakesh Shivanna}, \bibinfo{person}{Derek Cheng}, \bibinfo{person}{Sagar Jain}, \bibinfo{person}{Dong Lin}, \bibinfo{person}{Lichan Hong}, {and} \bibinfo{person}{Ed Chi}.} \bibinfo{year}{2021}\natexlab{}.
\newblock \showarticletitle{DCN V2: Improved Deep \& Cross Network and Practical Lessons for Web-Scale Learning to Rank Systems}. In \bibinfo{booktitle}{\emph{Proceedings of the Web Conference 2021}}. \bibinfo{pages}{1785–1797}.
\newblock


\bibitem[Wang et~al\mbox{.}(2024)]%
        {wang2024can}
\bibfield{author}{\bibinfo{person}{Yuling Wang}, \bibinfo{person}{Changxin Tian}, \bibinfo{person}{Binbin Hu}, \bibinfo{person}{Yanhua Yu}, \bibinfo{person}{Ziqi Liu}, \bibinfo{person}{Zhiqiang Zhang}, \bibinfo{person}{Jun Zhou}, \bibinfo{person}{Liang Pang}, {and} \bibinfo{person}{Xiao Wang}.} \bibinfo{year}{2024}\natexlab{}.
\newblock \showarticletitle{Can Small Language Models be Good Reasoners for Sequential Recommendation?}. In \bibinfo{booktitle}{\emph{Proceedings of the ACM on Web Conference 2024}}. \bibinfo{pages}{3876--3887}.
\newblock


\bibitem[Wu et~al\mbox{.}(2023)]%
        {wu2023survey}
\bibfield{author}{\bibinfo{person}{Likang Wu}, \bibinfo{person}{Zhi Zheng}, \bibinfo{person}{Zhaopeng Qiu}, \bibinfo{person}{Hao Wang}, \bibinfo{person}{Hongchao Gu}, \bibinfo{person}{Tingjia Shen}, \bibinfo{person}{Chuan Qin}, \bibinfo{person}{Chen Zhu}, \bibinfo{person}{Hengshu Zhu}, \bibinfo{person}{Qi Liu}, {et~al\mbox{.}}} \bibinfo{year}{2023}\natexlab{}.
\newblock \showarticletitle{A Survey on Large Language Models for Recommendation}.
\newblock \bibinfo{journal}{\emph{arXiv preprint arXiv:2305.19860}} (\bibinfo{year}{2023}).
\newblock


\bibitem[Xi et~al\mbox{.}(2024a)]%
        {xi2023towards}
\bibfield{author}{\bibinfo{person}{Yunjia Xi}, \bibinfo{person}{Weiwen Liu}, \bibinfo{person}{Jianghao Lin}, \bibinfo{person}{Xiaoling Cai}, \bibinfo{person}{Hong Zhu}, \bibinfo{person}{Jieming Zhu}, \bibinfo{person}{Bo Chen}, \bibinfo{person}{Ruiming Tang}, \bibinfo{person}{Weinan Zhang}, \bibinfo{person}{Rui Zhang}, {et~al\mbox{.}}} \bibinfo{year}{2024}\natexlab{a}.
\newblock \showarticletitle{Towards open-world recommendation with knowledge augmentation from large language models}. In \bibinfo{booktitle}{\emph{Proceedings of the ACM on Recommender Systems}}.
\newblock


\bibitem[Xi et~al\mbox{.}(2024b)]%
        {xi2024memocrs}
\bibfield{author}{\bibinfo{person}{Yunjia Xi}, \bibinfo{person}{Weiwen Liu}, \bibinfo{person}{Jianghao Lin}, \bibinfo{person}{Bo Chen}, \bibinfo{person}{Ruiming Tang}, \bibinfo{person}{Weinan Zhang}, {and} \bibinfo{person}{Yong Yu}.} \bibinfo{year}{2024}\natexlab{b}.
\newblock \showarticletitle{MemoCRS: Memory-enhanced Sequential Conversational Recommender Systems with Large Language Models}.
\newblock \bibinfo{journal}{\emph{arXiv preprint arXiv:2407.04960}} (\bibinfo{year}{2024}).
\newblock


\bibitem[Xi et~al\mbox{.}(2023)]%
        {xi2023device}
\bibfield{author}{\bibinfo{person}{Yunjia Xi}, \bibinfo{person}{Weiwen Liu}, \bibinfo{person}{Yang Wang}, \bibinfo{person}{Ruiming Tang}, \bibinfo{person}{Weinan Zhang}, \bibinfo{person}{Yue Zhu}, \bibinfo{person}{Rui Zhang}, {and} \bibinfo{person}{Yong Yu}.} \bibinfo{year}{2023}\natexlab{}.
\newblock \showarticletitle{On-device integrated re-ranking with heterogeneous behavior modeling}. In \bibinfo{booktitle}{\emph{Proceedings of the 29th ACM SIGKDD Conference on Knowledge Discovery and Data Mining}}. \bibinfo{pages}{5225--5236}.
\newblock


\bibitem[Xia et~al\mbox{.}(2023)]%
        {xia2023speculative}
\bibfield{author}{\bibinfo{person}{Heming Xia}, \bibinfo{person}{Tao Ge}, \bibinfo{person}{Peiyi Wang}, \bibinfo{person}{Si-Qing Chen}, \bibinfo{person}{Furu Wei}, {and} \bibinfo{person}{Zhifang Sui}.} \bibinfo{year}{2023}\natexlab{}.
\newblock \showarticletitle{Speculative decoding: Exploiting speculative execution for accelerating seq2seq generation}. In \bibinfo{booktitle}{\emph{Findings of the Association for Computational Linguistics: EMNLP 2023}}. \bibinfo{pages}{3909--3925}.
\newblock


\bibitem[Xia et~al\mbox{.}(2024)]%
        {xia2024unlocking}
\bibfield{author}{\bibinfo{person}{Heming Xia}, \bibinfo{person}{Zhe Yang}, \bibinfo{person}{Qingxiu Dong}, \bibinfo{person}{Peiyi Wang}, \bibinfo{person}{Yongqi Li}, \bibinfo{person}{Tao Ge}, \bibinfo{person}{Tianyu Liu}, \bibinfo{person}{Wenjie Li}, {and} \bibinfo{person}{Zhifang Sui}.} \bibinfo{year}{2024}\natexlab{}.
\newblock \showarticletitle{Unlocking efficiency in large language model inference: A comprehensive survey of speculative decoding}.
\newblock \bibinfo{journal}{\emph{arXiv preprint arXiv:2401.07851}} (\bibinfo{year}{2024}).
\newblock


\bibitem[Yang et~al\mbox{.}(2023)]%
        {yang2023predictive}
\bibfield{author}{\bibinfo{person}{Seongjun Yang}, \bibinfo{person}{Gibbeum Lee}, \bibinfo{person}{Jaewoong Cho}, \bibinfo{person}{Dimitris Papailiopoulos}, {and} \bibinfo{person}{Kangwook Lee}.} \bibinfo{year}{2023}\natexlab{}.
\newblock \showarticletitle{Predictive pipelined decoding: A compute-latency trade-off for exact LLM decoding}.
\newblock \bibinfo{journal}{\emph{arXiv preprint arXiv:2307.05908}} (\bibinfo{year}{2023}).
\newblock


\bibitem[Yu et~al\mbox{.}(2023)]%
        {yu2023self}
\bibfield{author}{\bibinfo{person}{Junliang Yu}, \bibinfo{person}{Hongzhi Yin}, \bibinfo{person}{Xin Xia}, \bibinfo{person}{Tong Chen}, \bibinfo{person}{Jundong Li}, {and} \bibinfo{person}{Zi Huang}.} \bibinfo{year}{2023}\natexlab{}.
\newblock \showarticletitle{Self-supervised learning for recommender systems: A survey}.
\newblock \bibinfo{journal}{\emph{IEEE Transactions on Knowledge and Data Engineering}} (\bibinfo{year}{2023}).
\newblock


\bibitem[Zhang et~al\mbox{.}(2023)]%
        {zhang2023collm}
\bibfield{author}{\bibinfo{person}{Yang Zhang}, \bibinfo{person}{Fuli Feng}, \bibinfo{person}{Jizhi Zhang}, \bibinfo{person}{Keqin Bao}, \bibinfo{person}{Qifan Wang}, {and} \bibinfo{person}{Xiangnan He}.} \bibinfo{year}{2023}\natexlab{}.
\newblock \showarticletitle{Collm: Integrating collaborative embeddings into large language models for recommendation}.
\newblock \bibinfo{journal}{\emph{arXiv preprint arXiv:2310.19488}} (\bibinfo{year}{2023}).
\newblock


\bibitem[Zhao et~al\mbox{.}(2023b)]%
        {zhao2023survey}
\bibfield{author}{\bibinfo{person}{Wayne~Xin Zhao}, \bibinfo{person}{Kun Zhou}, \bibinfo{person}{Junyi Li}, \bibinfo{person}{Tianyi Tang}, \bibinfo{person}{Xiaolei Wang}, \bibinfo{person}{Yupeng Hou}, \bibinfo{person}{Yingqian Min}, \bibinfo{person}{Beichen Zhang}, \bibinfo{person}{Junjie Zhang}, \bibinfo{person}{Zican Dong}, {et~al\mbox{.}}} \bibinfo{year}{2023}\natexlab{b}.
\newblock \showarticletitle{A survey of large language models}.
\newblock \bibinfo{journal}{\emph{arXiv preprint arXiv:2303.18223}} (\bibinfo{year}{2023}).
\newblock


\bibitem[Zhao et~al\mbox{.}(2023a)]%
        {zhao2023lookahead}
\bibfield{author}{\bibinfo{person}{Yao Zhao}, \bibinfo{person}{Zhitian Xie}, \bibinfo{person}{Chenyi Zhuang}, {and} \bibinfo{person}{Jinjie Gu}.} \bibinfo{year}{2023}\natexlab{a}.
\newblock \showarticletitle{Lookahead: An inference acceleration framework for large language model with lossless generation accuracy}.
\newblock \bibinfo{journal}{\emph{arXiv preprint arXiv:2312.12728}} (\bibinfo{year}{2023}).
\newblock


\bibitem[Zheng et~al\mbox{.}(2024b)]%
        {zheng2024adapting}
\bibfield{author}{\bibinfo{person}{Bowen Zheng}, \bibinfo{person}{Yupeng Hou}, \bibinfo{person}{Hongyu Lu}, \bibinfo{person}{Yu Chen}, \bibinfo{person}{Wayne~Xin Zhao}, \bibinfo{person}{Ming Chen}, {and} \bibinfo{person}{Ji-Rong Wen}.} \bibinfo{year}{2024}\natexlab{b}.
\newblock \showarticletitle{Adapting large language models by integrating collaborative semantics for recommendation}. In \bibinfo{booktitle}{\emph{2024 IEEE 40th International Conference on Data Engineering (ICDE)}}. IEEE, \bibinfo{pages}{1435--1448}.
\newblock


\bibitem[Zheng et~al\mbox{.}(2023)]%
        {zheng2023judging}
\bibfield{author}{\bibinfo{person}{Lianmin Zheng}, \bibinfo{person}{Wei-Lin Chiang}, \bibinfo{person}{Ying Sheng}, \bibinfo{person}{Siyuan Zhuang}, \bibinfo{person}{Zhanghao Wu}, \bibinfo{person}{Yonghao Zhuang}, \bibinfo{person}{Zi Lin}, \bibinfo{person}{Zhuohan Li}, \bibinfo{person}{Dacheng Li}, \bibinfo{person}{Eric Xing}, {et~al\mbox{.}}} \bibinfo{year}{2023}\natexlab{}.
\newblock \showarticletitle{Judging llm-as-a-judge with mt-bench and chatbot arena}.
\newblock \bibinfo{journal}{\emph{Advances in Neural Information Processing Systems}}  \bibinfo{volume}{36} (\bibinfo{year}{2023}), \bibinfo{pages}{46595--46623}.
\newblock


\bibitem[Zheng et~al\mbox{.}(2024a)]%
        {zheng2024harnessing}
\bibfield{author}{\bibinfo{person}{Zhi Zheng}, \bibinfo{person}{Wenshuo Chao}, \bibinfo{person}{Zhaopeng Qiu}, \bibinfo{person}{Hengshu Zhu}, {and} \bibinfo{person}{Hui Xiong}.} \bibinfo{year}{2024}\natexlab{a}.
\newblock \showarticletitle{Harnessing large language models for text-rich sequential recommendation}. In \bibinfo{booktitle}{\emph{Proceedings of the ACM on Web Conference 2024}}. \bibinfo{pages}{3207--3216}.
\newblock


\bibitem[Zhou et~al\mbox{.}(2018)]%
        {DIN}
\bibfield{author}{\bibinfo{person}{Guorui Zhou}, \bibinfo{person}{Xiaoqiang Zhu}, \bibinfo{person}{Chenru Song}, \bibinfo{person}{Ying Fan}, \bibinfo{person}{Han Zhu}, \bibinfo{person}{Xiao Ma}, \bibinfo{person}{Yanghui Yan}, \bibinfo{person}{Junqi Jin}, \bibinfo{person}{Han Li}, {and} \bibinfo{person}{Kun Gai}.} \bibinfo{year}{2018}\natexlab{}.
\newblock \showarticletitle{Deep Interest Network for Click-Through Rate Prediction}. In \bibinfo{booktitle}{\emph{Proceedings of the 24th ACM SIGKDD International Conference on Knowledge Discovery and Data Mining}}. \bibinfo{pages}{1059–1068}.
\newblock


\bibitem[Zhou et~al\mbox{.}(2023)]%
        {zhou2023distillspec}
\bibfield{author}{\bibinfo{person}{Yongchao Zhou}, \bibinfo{person}{Kaifeng Lyu}, \bibinfo{person}{Ankit~Singh Rawat}, \bibinfo{person}{Aditya~Krishna Menon}, \bibinfo{person}{Afshin Rostamizadeh}, \bibinfo{person}{Sanjiv Kumar}, \bibinfo{person}{Jean-Fran{\c{c}}ois Kagy}, {and} \bibinfo{person}{Rishabh Agarwal}.} \bibinfo{year}{2023}\natexlab{}.
\newblock \showarticletitle{Distillspec: Improving speculative decoding via knowledge distillation}.
\newblock \bibinfo{journal}{\emph{arXiv preprint arXiv:2310.08461}} (\bibinfo{year}{2023}).
\newblock


\bibitem[Zhu et~al\mbox{.}(2024)]%
        {zhu2024collaborative}
\bibfield{author}{\bibinfo{person}{Yaochen Zhu}, \bibinfo{person}{Liang Wu}, \bibinfo{person}{Qi Guo}, \bibinfo{person}{Liangjie Hong}, {and} \bibinfo{person}{Jundong Li}.} \bibinfo{year}{2024}\natexlab{}.
\newblock \showarticletitle{Collaborative large language model for recommender systems}. In \bibinfo{booktitle}{\emph{Proceedings of the ACM on Web Conference 2024}}. \bibinfo{pages}{3162--3172}.
\newblock


\bibitem[Zhu et~al\mbox{.}(2023)]%
        {zhu2023large}
\bibfield{author}{\bibinfo{person}{Yutao Zhu}, \bibinfo{person}{Huaying Yuan}, \bibinfo{person}{Shuting Wang}, \bibinfo{person}{Jiongnan Liu}, \bibinfo{person}{Wenhan Liu}, \bibinfo{person}{Chenlong Deng}, \bibinfo{person}{Zhicheng Dou}, {and} \bibinfo{person}{Ji-Rong Wen}.} \bibinfo{year}{2023}\natexlab{}.
\newblock \showarticletitle{Large language models for information retrieval: A survey}.
\newblock \bibinfo{journal}{\emph{arXiv preprint arXiv:2308.07107}} (\bibinfo{year}{2023}).
\newblock


\end{thebibliography}

\end{document}